\documentclass[fleqn,usenatbib]{mnras}

\usepackage{newtxtext,newtxmath}
\usepackage[T1]{fontenc}

\usepackage{graphicx}
\usepackage{amsmath}
\usepackage{ulem}

\newcommand{\llangle}{\langle\!\langle}
\newcommand{\rrangle}{\rangle\!\rangle}

\title[Stationarity and angular momentum conservation in pulsar spin noise]{Stationarity and angular momentum conservation in pulsar spin noise}

\author[Reginald Christian Bernardo]{
Reginald Christian Bernardo,\thanks{E-mail: reginald.christian.bernardo@aei.mpg.de}
\\
Max Planck Institute for Gravitational Physics (Albert Einstein Institute), \\ Leibniz Universit\"{a}t Hannover, Callinstrasse 38, D-30167 Hannover, Germany \\
}

\date{Accepted XXX. Received YYY; in original form ZZZ}
\pubyear{\the\year{}}

\begin{document}
\label{firstpage}
\pagerange{\pageref{firstpage}--\pageref{lastpage}}
\maketitle

\begin{abstract}
We present an analysis of pulsar spin noise based on physically-motivated two-component models of spin wandering. We focus on two models, distinguished by their total angular momentum dynamics: a singular, nonstationary model with a diffusive total angular momentum, and a minimal, stationary model anchored on a conserved total angular momentum. We develop scalable Gaussian process methods and, using mock data, show that the two models are fully testable and distinguishable with full-state observations, i.e., simultaneous independent data on the crust and the superfluid rotational states. This paves a path to testing stationarity in pulsar spin noise, potentially achievable with joint continuous gravitational wave observations and radio timing of pulsars. However, robust inferences and predictions are harder to achieve, and depend on the priors and data quality, when only one component is observationally accessible.
\end{abstract}


\section{Introduction}
\label{sec:introduction}

Pulsars are highly magnetized, stably rotating neutron stars that send beams of electromagnetic radiation at regular intervals \citep{Hewish:1968bj, Pilkington:1968bk, Wallace:1977he, Taylor:1977wj}. For this reason, they are regarded as one of the Universe's most precise natural clocks \citep{Lorimer:2001vd, Kaspi:2016jkv, Bassa:2017zpe, Fermi-LAT:2018jvx, Nieder:2020yqy, Clark:2025lai, Bagchi:2025hnw}. This exquisite timing precision has been used to set constraints on fundamental physics through time delays caused by fields such as strong gravity \citep{Manchester:2015mda, Kramer:2021jcw, Freire:2024adf}, dark matter \citep{Khmelnitsky:2013lxt, Porayko:2014rfa, Porayko:2018sfa, EuropeanPulsarTimingArray:2023egv}, and gravitational waves \citep{Hulse:1974eb, 1982ApJ...253..908T, Hulse:1994zz, Damour:2014tpa}. Pulsar timing arrays have made the best out of this to search for a gravitational wave background signal via spatially-correlated time delays \citep{NANOGrav:2023gor, EPTA:2023fyk, Reardon:2023gzh, Xu:2023wog}.

The precision achievable with pulsar timing has a limit that remains incompletely understood. Once a deterministic timing model -- accounting for the pulsar's spin period, period derivative, orbital dynamics, and other deterministic effects -- is subtracted, glitches, timing noise, mode switching, and nulling are among the phenomena that set the ceiling on a pulsar's utility for testing fundamental physics through time delays. Physical pictures exist for some of these. Glitches are reasonably well described by a two-step pinning and unpinning process of superfluid vortices entrained in the neutron star's crust \citep{1969Natur.224..872B, 2012PhRvL.109x1103A, 2013PhRvL.110a1101C}; timing noise is commonly depicted as a red stochastic process in the neutron star's spin or angular momentum, i.e., spin wandering \citep{1975ApJS...29..453G, 2010MNRAS.402.1027H}; and mode switching and nulling are associated with the dynamics of the magnetosphere \citep{1970Natur.228...42B, 1982ApJ...258..776B, 2010MNRAS.408L..41T}.
However, a unified physical picture connecting these phenomena, and in particular tying timing noise to a falsifiable physical origin, remains lacking.

We start with glitches and timing noise, both of which have been given phenomenological accounts within a single body-averaged two-component dynamical model of a neutron star's rotational degrees of freedom. A glitch is preceded by the build-up of a crust-core lag and an angular momentum reservoir that is released in a sudden spin-up of the crust \citep{2020A&A...642A.223M, 2012PhRvL.109x1103A, 2013PhRvL.110a1101C}. Timing noise, on the other hand, is associated with stochastic torques that continuously act on the components of the neutron star. However, the two-component models considered so far to explain radio timing noise correspond to a nonstationary Gaussian process \citep{Meyers:2021myb,Meyers:2021slh,Antonelli:2022gqw,Bernardo:2026opv}, whose nonstationarity can be traced to a diffusive total angular momentum mode excited by the stochastic torques. A stationary process can instead be motivated by imposing that the total angular momentum is conserved, or that stochastic torques compel only internal angular momentum exchanges \citep{Antonelli:2022gqw, Antonelli:2025vhw}.
Conservation of total angular momentum therefore sits at the heart of the question of whether pulsar spin noise is stationary, and hints at a potentially falsifiable prediction of pulsar physics: testing stationarity in spin noise amounts to testing whether the total angular momentum of a neutron star is conserved. This work sets out to develop methods geared toward testing stationarity in pulsar spin noise.

However, there is an observational caveat. Radio pulsar timing measures only the crust; in principle, we take this to be the part of the star that is strongly electromagnetically-coupled. The detected pulses, and our understanding of their emission, are tied to the crust's rotational state. Theory offers insights into the star's interior dynamics \citep{Prix:1999gk, Prix:2001xc, Prix:2002jn, Andersson:2005rs, Melatos:2013rca, Haskell:2015jra, Haskell:2017lkl}, and continuous gravitational waves may complement radio timing by probing the star's interior rotation
\citep{Jaranowski:1998qm, 2009ASSL..357..651P, 2009astro2010S.229O, 2021arXiv210403137H, Ming:2021xtz, Dergachev:2025hwp, Dergachev:2025ead, 2026arXiv260718219M},
completing the picture beyond the radio band. This raises a question that this work takes up directly: if only a single axis of a multidimensional rotational dynamics is observed, what can we robustly say about the underlying model, or about the interior state it predicts?

Previous studies have focused on parameter identifiability. Electromagnetic timing data alone can only recover a subset of the model parameters, and gravitational wave data on the superfluid rotation is needed to break the remaining degeneracies \citep{Meyers:2021myb,Meyers:2021slh}. The model's power spectral density has been used to extract analytic timing-noise-strength scalings \citep{Antonelli:2022gqw}, and future joint electromagnetic-gravitational wave correlations have been proposed to observationally distinguish internal and magnetospheric noise \citep{Antonelli:2025vhw}.

Robust tests of these physical models must also confront the practical realities of pulsar timing data \citep{2014ApJ...788...33K, 2015ITPAM..38..252A, 2015MNRAS.446.1170V, Goncharov:2019mei}: decades-long, unevenly sampled time series with heteroscedastic measurement uncertainties, frequently analyzed jointly across many pulsars in a pulsar timing array. A direct Gaussian process likelihood evaluation scales as ${\cal O}(N^3)$ in the number of observations, which becomes prohibitive as data spans lengthen and as more pulsars, or additional rotational-state observables, enter a joint analysis. This motivates the development of scalable, physically-motivated Gaussian process methods \citep{2018PhRvE..98a2136S, 2017AJ....154..220F} that retain the ${\cal O}(N)$ efficiency of phenomenological kernels while remaining anchored to the dynamics of spin wandering.

We first revisit the two-component model of spin wandering (Section \ref{sec:spin_wandering}) and its analytical solutions (Section \ref{sec:analytical_solutions}), which together provide the foundation for constraining stationarity in spin noise from timing observations. Building on this foundation, we develop scalable, physically-motivated Gaussian process methods for pulsar spin noise using well-established diffusion theory and the corresponding analytical solutions (Section \ref{sec:fast_bayesian}). We then test the method on mock angular velocity residuals generated from both the nonstationary and the stationary spin noise models (Sections \ref{sec:singular_simulations}--\ref{sec:minimal_simulations}). We conclude with a discussion of future work (Section \ref{sec:conclusions}).

We include four appendices with additional details. Appendix \ref{sec:two_fluid} motivates the two-component model starting from a two-fluid description of a neutron star; Appendix \ref{sec:fluctuation_dissipation_theorem} briefly reviews the fluctuation-dissipation theorem; Appendix \ref{sec:2d_brownian} illustrates the fast Bayesian method of Section \ref{sec:fast_bayesian} on 2D Brownian motion; and Appendix \ref{sec:full_state_semiseparable} presents the full-state semiseparable likelihood for the two-component model.

\section{Spin wandering with two components}
\label{sec:spin_wandering}

Consider the neutron star two-component model with spin wandering described by the equations of motion \citep{1969Natur.224..872B, 1975ApJS...29..453G, Meyers:2021myb,Meyers:2021slh, Antonelli:2022gqw}
\begin{align}
    \label{eq:eom_crust}
    I_\text{c} \dot{\Omega}_\text{c} &= -\dfrac{I_{\rm c}}{\tau_{\rm c}} \left( \Omega_{\rm c} - \Omega_{\rm s} \right) + N_{\rm c} + \xi_{\rm c}(t) \\
    \label{eq:eom_superfluid}
    I_\text{s} \dot{\Omega}_\text{s} &= \dfrac{I_{\rm s}}{\tau_{\rm s}} \left( \Omega_{\rm c} - \Omega_{\rm s} \right) + N_{\rm s} + \xi_{\rm s}(t)
\end{align}
where subscripts c and s correspond to crust and superfluid components, respectively, $(I_\text{c}, I_\text{s})$ are constant moments of inertia, $(\Omega_\text{c}, \Omega_\text{s})$ are angular velocities, and $(N_\text{c}, N_\text{s})$ are external torques. The time scales $\tau_{\rm c}$ and $\tau_{\rm s}$ characterize nondissipative mutual friction between the crust and the superfluid, and the stochastic torques $(\xi_{\rm c}(t), \xi_{\rm s}(t))$ represent the random fluctuations. Mutual friction terms balance each other in the limit $I_{\rm s}/I_{\rm c}=\tau_{\rm s}/\tau_{\rm c}$.
We adopt this detailed-balance regime throughout; its consequence for the total angular momentum equation of motion is made explicit below.
The stochastic torques $\xi_{\rm c}(t)$ and $\xi_{\rm s}(t)$ are assumed to be Gaussian white noise processes with zero mean and correlation functions
\begin{align}
    \label{eq:xic_xic}
    \langle \xi_{\rm c}(t) \xi_{\rm c}(t') \rangle &= Q_{\rm c} I_{\rm c}^2 \delta(t - t') \\
    \label{eq:xis_xis}
    \langle \xi_{\rm s}(t) \xi_{\rm s}(t') \rangle &= Q_{\rm s} I_{\rm s}^2 \delta(t - t') \\
    \label{eq:xic_xis}
    \langle \xi_{\rm c}(t) \xi_{\rm s}(t') \rangle &= 0    
\end{align}
where $(Q_{\rm c}, Q_{\rm s})$ are diffusion coefficients. The notation $\langle \cdot \rangle$ denotes an ensemble average; or the average over a large number of realizations of a stochastic process. Appendix \ref{sec:two_fluid} motivates (\ref{eq:eom_crust}-\ref{eq:xic_xis}) by coarse-graining a two-fluid model.

The model (\ref{eq:eom_crust}-\ref{eq:xic_xis}) has six free parameters -- $(\tau_{\rm c}, \tau_{\rm s}, N_{\rm c} / I_{\rm c}, N_{\rm s} / I_{\rm s}, Q_{\rm c}, Q_{\rm s})$, a relaxation time scale, a diffusion coefficient, and an external torque coefficient for each component. The time scales can also be expressed in terms of a crust-core coupling time scale $\tau$ and ratio $r$ given by
\begin{align}
    \label{eq:tau_def}
    \tau &= \dfrac{\tau_{\rm c} \tau_{\rm s}}{\tau_{\rm c} + \tau_{\rm s}} \\
    \label{eq:r_def}
    r &= \dfrac{\tau_{\rm s}}{\tau_{\rm c}} \,.
\end{align}
For a random variable, $O(t)$, with an ensemble average that approaches a finite constant limit at large times compared to the system relaxation time, we define constant steady state values $\llangle \cdot \rrangle$ by
\begin{equation}
    \llangle O(t) \rrangle \equiv \lim_{t\rightarrow \infty} \langle O(t) \rangle \equiv \langle O(t) \rangle_\infty \,.
\end{equation}
The constant external torques weighted by the moments of inertia can be expressed in terms of the constant steady-state crust angular acceleration, $\llangle \alpha_{\rm c} \rrangle = \langle \dot{\Omega}_{\rm c} \rangle_{\infty}$, and crust-core spin lag, $\llangle \Delta \Omega \rrangle = \langle {\Omega}_{\rm c} - {\Omega}_{\rm s} \rangle_{\infty}$, through the following equations:
\begin{align}
    \label{eq:lag_ss}
    \llangle \Delta \Omega \rrangle &= \tau \left( \dfrac{N_{\rm c}}{I_{\rm c}} - \dfrac{N_{\rm s}}{I_{\rm s}} \right) \\
    \label{eq:alpha_ss}
    \llangle \alpha_{\rm c} \rrangle &= \dfrac{1}{\tau_{\rm c} + \tau_{\rm s}} \left( \dfrac{\tau_{\rm c} N_{\rm c}}{I_{\rm c}} + \dfrac{\tau_{\rm s} N_{\rm s}}{I_{\rm s}} \right) \,.
\end{align}
The processes $\dot{\Omega}_{\rm c}$ and $\Omega_{\rm c} - \Omega_{\rm s}$ have constant steady state values and are stationary (\cite{Meyers:2021myb,Meyers:2021slh,Bernardo:2026opv}). (\ref{eq:tau_def}-\ref{eq:lag_ss}) can be inverted to obtain:
\begin{align}
    \tau_{\rm c} &= \tau \left( 1 + r^{-1} \right) \\
    \tau_{\rm s} &= \tau \left( 1 + r \right) \\
    \dfrac{N_{\rm c}}{I_{\rm c}} &= \llangle \alpha_{\rm c} \rrangle + \dfrac{\llangle \Delta \Omega \rrangle}{\tau} \dfrac{r}{1 + r} \\
    \dfrac{N_{\rm s}}{I_{\rm s}} &= \llangle \alpha_{\rm c} \rrangle - \dfrac{\llangle \Delta \Omega \rrangle}{\tau} \dfrac{1}{1 + r} \,.
\end{align}
Then, the free parameters of the two-component model with spin wandering can be treated as $(\tau, r, \llangle \alpha_{\rm c} \rrangle, \llangle \Delta \Omega \rrangle, Q_{\rm c}, Q_{\rm s})$. \cite{ONeill:2024uiw, Dong:2025nho} have shown that four out of six of the parameters of the model ($\tau, \llangle \Delta \alpha_{\rm c} \rrangle, Q_{\rm c}, Q_{\rm s}$) can be constrained by electromagnetic observations of the pulsar phase; $\llangle \Delta \alpha_{\rm c} \rrangle$ can be associated with the spin down, or more precisely the spin down rate at steady state, and $(\tau, Q_{\rm c}, Q_{\rm s})$ to timing noise exhibited through spin wandering.

We shall refer to $(\tau_{\rm c}, \tau_{\rm s}, N_{\rm c} / I_{\rm c}, N_{\rm s} / I_{\rm s}, Q_{\rm c}, Q_{\rm s})$ as component parametrization and $(\tau, r, \llangle \alpha_{\rm c} \rrangle, \llangle \Delta \Omega \rrangle, Q_{\rm c}, Q_{\rm s})$ as eigenbasis parametrization, respectively. For this work, we will work directly with the angular velocities $\Omega_{\rm i}(t)$'s or spin frequencies $f_{\rm i}(t) = \Omega_{\rm i}(t)/(2\pi)$ and their residuals \citep{ONeill:2024uiw}.\footnote{
The pulsar phase $\phi(t)$ is related to
$\Omega_{\rm c}(t)$ through 
$    \phi(t) = \int_0^t dt' \, \Omega_{\rm c}(t')$ \citep{ONeill:2024uiw, Dong:2025nho}.
This is related to the timing residual $r(t)$ via $r(t) \simeq \delta \phi / ( 2\pi\nu )$ where $\nu$ is a nominal spin frequency and $\delta \phi = \phi - \phi^{\rm M}$ is the observed phase residual relative to a particular timing model $\phi^{\rm M}(t)$ that factors in spin-down and other deterministic effects.
}

\subsection*{Spin noise and the conservation of total angular momentum}

The model (\ref{eq:eom_crust}-\ref{eq:xic_xis}) can be parametrized with the crust-core lag $\Omega_-$ and the total angular momentum $J$:
\begin{align}
    \Omega_- \equiv \, & \Omega_{\rm c} - \Omega_{\rm s} \\
    J \equiv \, & I_{\rm c} \Omega_{\rm c} + I_{\rm s} \Omega_{\rm s} \,.
\end{align}
These system eigenmodes can be identified by diagonalizing the equations of motion \citep{Meyers:2021myb,Meyers:2021slh, Antonelli:2022gqw,Bernardo:2026opv} and lead to the dynamical system
\begin{align}
    \dot{J} &= N_{\rm c} + N_{\rm s} + \xi_{\rm c}(t) + \xi_{\rm s}(t) \\
    \dot{\Omega}_- &= - \dfrac{\Omega_-}{\tau} +  \dfrac{N_{\rm c}}{I_{\rm c}} - \dfrac{N_{\rm s}}{I_{\rm s}} + \dfrac{\xi_{\rm c}(t)}{I_{\rm c}} - \dfrac{\xi_{\rm s}(t)}{I_{\rm s}} \,.
\end{align}
This simplified form of $\dot J$ follows from the detailed-balance condition $I_{\rm s}/I_{\rm c}=\tau_{\rm s}/\tau_{\rm c}$ imposed above. Without it, diagonalizing (\ref{eq:eom_crust}-\ref{eq:eom_superfluid}) instead gives $\dot J = \left(I_{\rm s}/\tau_{\rm s} - I_{\rm c}/\tau_{\rm c}\right)\Omega_- + N_{\rm c}+N_{\rm s}+\xi_{\rm c}(t)+\xi_{\rm s}(t)$, so total angular momentum is sourced by the crust-core lag whenever detailed balance is broken.
A physical conclusion can be reached with the above parametrization: even in the absence of deterministic external torques, stochastic torques will continue to pump in angular momentum into the system. The crust-core lag is an OU process. This turns the process nonstationary as a linear combination of a diffusive total angular momentum eigenmode stationary crust-core lag eigenmode. For this reason, we refer to the model (\ref{eq:eom_crust}-\ref{eq:xic_xis}) as singular, nonstationary, or both in what follows. The word `singular' is used to describe the general two-component model that is described by a singular drift matrix (Section \ref{sec:analytical_solutions}).

A minimal physical model that conserves angular momentum can be explicitly constructed by setting $\xi_{\rm c}(t) \equiv -\xi_{\rm s}(t)$ \citep{Meyers:2021myb}. This can be realized by imposing that the total angular momentum mode is internally conserved \citep{Antonelli:2022gqw, Antonelli:2025vhw}, or by invoking the fluctuation-dissipation theorem (Appendix \ref{sec:fluctuation_dissipation_theorem}).
An appealing feature of this model is that it has one parameter $N_{\rm c}/I_{\rm c}$ that can be associated to spin-down, and three free parameters $(\tau, x_{\rm s} = r/(1+r), Q)$ to spin wandering;
where $x_{\rm c}= I_{\rm c}/(I_{\rm c} + I_{\rm s})$ and $x_{\rm s} = 1 - x_{\rm c}$ are fractional moments of inertia in the crust and the superfluid.
The eigenbasis equations of motion of this model are given by
\begin{align}
\label{eq:Jeom_minimal}
    \dot{J} &= N_{\rm c} \\
\label{eq:lageom_minimal}
    \dot{\Omega}_- &= - \dfrac{\Omega_-}{\tau} +  \dfrac{N_{\rm c}}{I_{\rm c}} + \dfrac{ \xi(t) }{ x_{\rm c} x_{\rm s} I } \,,
\end{align}
where $\xi(t) = -\xi_{\rm c}(t) = \xi_{\rm s}(t)$, and
\begin{equation}
\label{eq:wncov_minimal}
    \langle \xi(t) \xi(t') \rangle = Q I^2 \delta( t - t' ) \,.
\end{equation}
The component form of the equations of motion is given by \citep{Antonelli:2022gqw, Antonelli:2025vhw}
\begin{align}
    \label{eq:eom_crust_minimal}
    I_\text{c} \dot{\Omega}_\text{c} &= -\dfrac{ I_{\rm c} I_{\rm s} }{I_{\rm c} + I_{\rm s} } \dfrac{\left( \Omega_{\rm c} - \Omega_{\rm s} \right)}{\tau} + N_{\rm c} + \xi(t) \\
    \label{eq:eom_superfluid_minimal}
    I_\text{s} \dot{\Omega}_\text{s} &= \dfrac{ I_{\rm c} I_{\rm s} }{I_{\rm c} + I_{\rm s} } \dfrac{\left( \Omega_{\rm c} - \Omega_{\rm s} \right)}{\tau} - \xi(t) \,.
\end{align}
The above equations reveal that nondissipative mutual friction is expressed symmetrically between the two components.
Most importantly, this model accommodates a single random dynamical variable, $\Omega_-$, that is an OU process, and is therefore inherently 1D and stationary. Therefore, we refer to the model (\ref{eq:Jeom_minimal}-\ref{eq:wncov_minimal}) as minimal, stationary or both. 'Minimal' is used relatively to indicate a model with fewer parameters and dimensions.

\begin{figure*}
    \centering
    \includegraphics[width=0.9\textwidth]{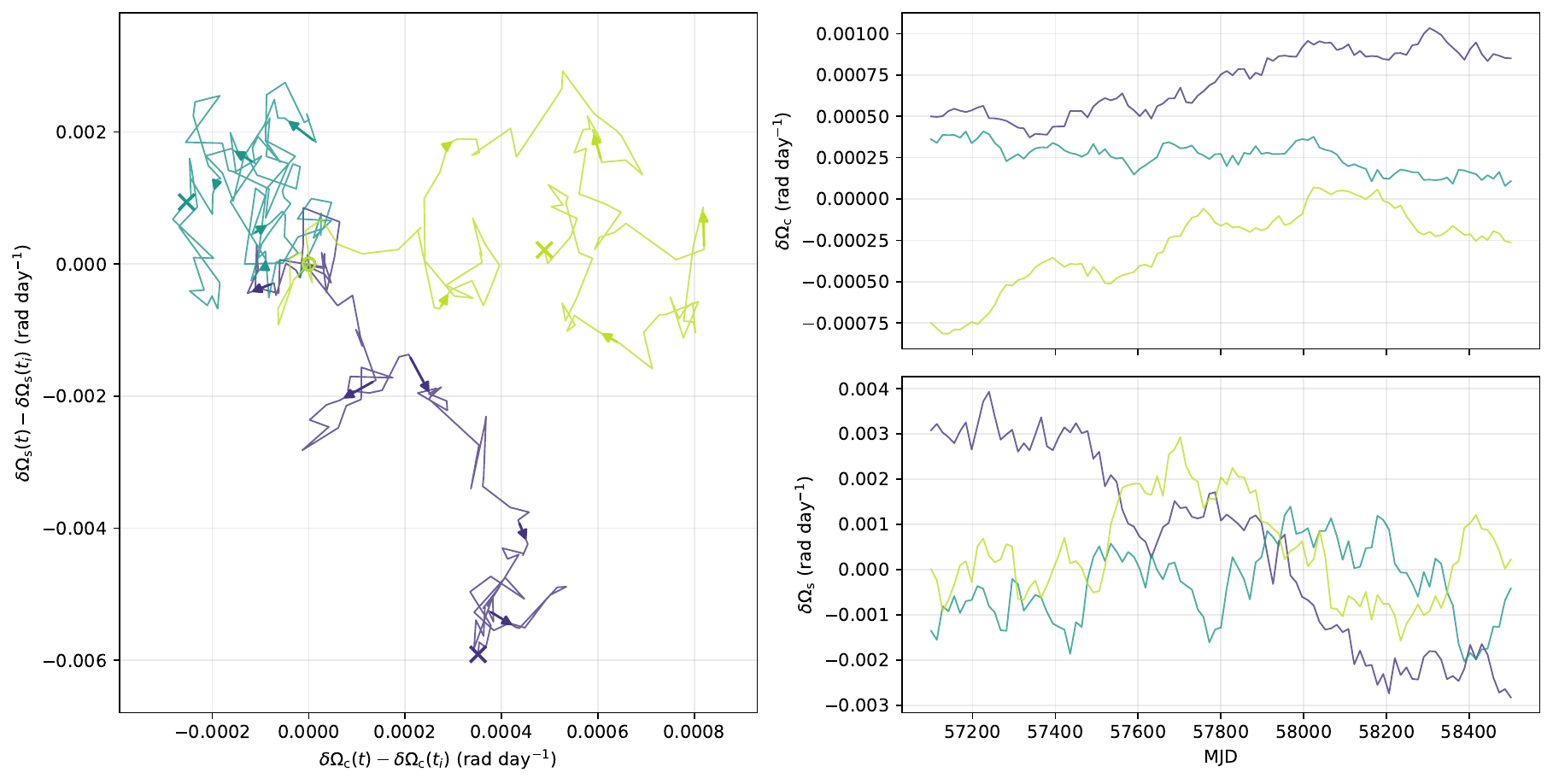}
    \includegraphics[width=0.9\textwidth]{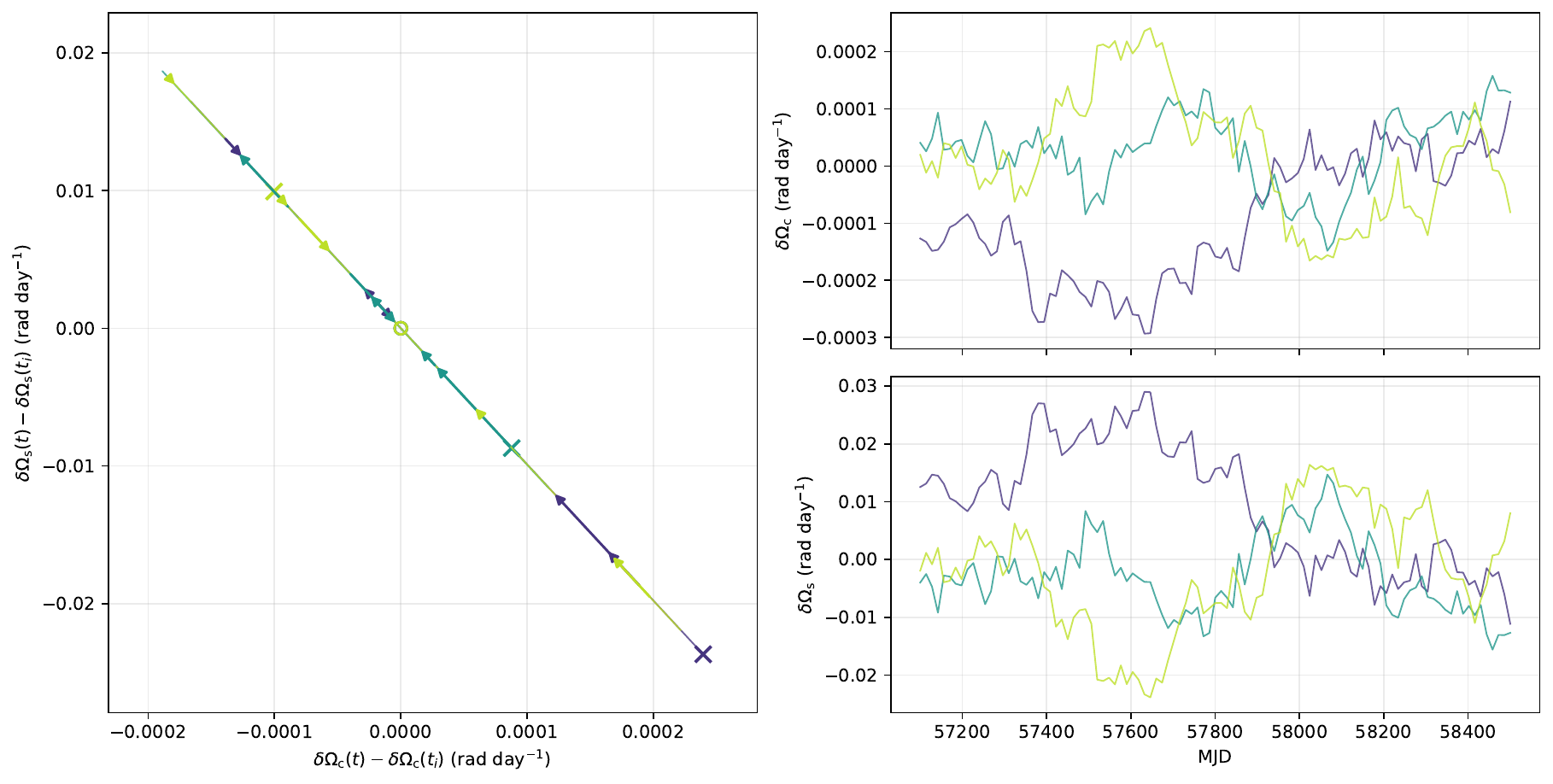}
    \caption{Sample realizations of the singular, nonstationary model (\ref{eq:eom_crust}--\ref{eq:xic_xis}) and the minimal, stationary model (\ref{eq:Jeom_minimal}-\ref{eq:wncov_minimal}) with parameters in Tables \ref{tab:singular_parameters} and \ref{tab:minimal_parameters}.}
    \label{fig:sample_paths}
\end{figure*}

Realizations of the singular-nonstationary (\ref{eq:eom_crust}-\ref{eq:xic_xis}) and minimal-stationary (\ref{eq:Jeom_minimal}-\ref{eq:wncov_minimal}) are shown in Figure \ref{fig:sample_paths}.

\section{Analytical solutions}
\label{sec:analytical_solutions}

The model (\ref{eq:eom_crust}-\ref{eq:xic_xis}) can be written as a shifted bivariate OU process:\footnote{`Shifted' mathematically means that the process is of nonzero mean, or physically that external forces act on the system.}
\begin{equation}
\label{eq:eom_2cm_matrix}
    \dot{\boldsymbol{\Omega}} = -\mathbf{D}\boldsymbol{\Omega} + \mathbf{N} + \boldsymbol{\sigma}\dot{\mathbf{W}}
\end{equation}
where
\begin{equation}
    \boldsymbol{\Omega} = \begin{bmatrix} \Omega_{\rm c} \\ \Omega_{\rm s} \end{bmatrix} \,,
\end{equation}
\begin{equation}
\label{eq:drift_matrix_2cm}
    \mathbf{D} = \begin{bmatrix} 1/\tau_{\rm c} & -1/\tau_{\rm c} \\ -1/\tau_{\rm s} & 1/\tau_{\rm s} \end{bmatrix} \,,
\end{equation}
\begin{equation}
\label{eq:torque_matrix_2cm}
    \mathbf{N} = \begin{bmatrix} N_{\rm c}/I_{\rm c} \\ N_{\rm s}/I_{\rm s} \end{bmatrix} \,,
\end{equation}
\begin{equation}
\label{eq:sigma_matrix}
    \boldsymbol{\sigma} = \begin{bmatrix} \sqrt{Q_{\rm c}} & 0 \\ 0 & \sqrt{Q_{\rm s}}\end{bmatrix} \,,
\end{equation}
and $\dot{\mathbf{W}}(t)$ is a Gaussian white noise process with zero mean and correlation function
\begin{equation}
\label{eq:wn_cov_2cm_matrix}
    \langle \dot{W}_i(t) \dot{W}_j(t') \rangle = \delta_{ij} \delta(t - t') \,.
\end{equation}
We refer to $\mathbf{D}$ and $\boldsymbol{\sigma}$ as the drift and volatility matrices, respectively. The volatility matrix gives rise to the diffusion matrix or process covariance:
\begin{equation}
\label{eq:sigma_matrix_2cm}
    \mathbf{Q} = \boldsymbol{\sigma} \boldsymbol{\sigma}^\top \,.
\end{equation}
The matrix $\mathbf{Q}$ is a symmetric positive semi-definite matrix with diffusion coefficients. The off-diagonal elements of 
$\mathbf{Q}$ encode correlations between the stochastic torques acting on the crust and superfluid components; the model (\ref{eq:eom_crust}-\ref{eq:xic_xis}) assumes these are uncorrelated.
The eigenvalues of
$\mathbf{D}$ are given by $\lambda_{\mathbf{D}} = \tau^{-1}, 0$, and so 
$\mathbf{D}$ is singular, i.e., $|\mathbf{D} |= 0$. The first eigenvalue belongs to a stationary eigenmode or damped random walk with a finite relaxation time $\tau$. The second corresponds to a nonstationary eigenmode or pure random walk.
The former can be identified with the crust-core lag and the latter to the total angular momentum mode \citep{Bernardo:2026opv}. This implies that the Gaussian process $\boldsymbol{\Omega}$ is nonstationary and not mean-reverting (\cite{Meyers:2021myb,Meyers:2021slh,Antonelli:2022gqw,Bernardo:2026opv}).

This implies that the analytic solution to the multivariate OU process expressed in terms of the solution of the Lyapunov equation is not directly applicable \citep{2018PhRvE..98a2136S}.
Analytical (time-domain, dynamical) solutions to (\ref{eq:eom_crust}-\ref{eq:xic_xis}) were presented in \cite{Bernardo:2026opv} to highlight the process' nonstationary behavior -- that can be traced to stochastic variations in the total angular momentum.

We circle around this limitation by regularizing the drift matrix as follows. In place of $\mathbf{D}$ in \eqref{eq:eom_2cm_matrix}, we use $\mathbf{D}_\eta$ given by
\begin{equation}
\label{eq:drift_matrix_2cm_reg}
    \mathbf{D}_\eta = \begin{bmatrix} 1/\tau_{\rm c} & -1/\tau_{\rm c} + 1/\eta \\ -1/\tau_{\rm s} + 1/\eta & 1/\tau_{\rm s} \end{bmatrix} \,,
\end{equation}
where $\eta > 0$ is a constant regularization parameter. The time scale $\eta$ controls the growth of the total angular momentum in the regularized model. The eigenvalues of $\mathbf{D}_\eta$ are given by $\lambda = \tau^{-1} - \eta^{-1}$ and $\lambda = \eta^{-1}$, and so
$\mathbf{D}_\eta$
has strictly positive eigenvalues in the parameter space $\eta > \tau > 0$. For a relaxation time $\tau = {\cal O}(1)\,{\rm yr}$ and an observation window $T_{\rm obs} = {\cal O}(1)\,{\rm yr}$, an appropriate choice of $\eta$ is $\eta \gg \tau, T_{\rm obs}$.\footnote{The original singular process can be approached informally by taking increasingly large values of the regularization parameter, or formally in the limit $\eta \to \infty$ of the regularized process.} Then, the standard OU analytical solution can be applied to the regularized process \citep{2018PhRvE..98a2136S}.

The solution to \eqref{eq:eom_2cm_matrix} can be presented as follows.
We write down
\begin{equation}
\label{eq:omega_decomp}
    \boldsymbol{\Omega}=\boldsymbol{\Omega}_{\rm D} + \boldsymbol{\Omega}_{\rm S}
\end{equation}
where
\begin{equation}
\label{eq:eom_det}
    \dot{\boldsymbol{\Omega}}_{\rm D} = -\mathbf{D}_\eta \boldsymbol{\Omega}_{\rm D} + \mathbf{N}
\end{equation}
and
\begin{equation}
\label{eq:eom_stoc}
    \dot{\boldsymbol{\Omega}}_{\rm S} = -\mathbf{D}_\eta \boldsymbol{\Omega}_{\rm S} + \boldsymbol{\sigma}\dot{\mathbf{W}} \,.
\end{equation}
The linearity of \eqref{eq:eom_2cm_matrix} guarantees that it is satisfied by \eqref{eq:omega_decomp} together with deterministic and stochastic inhomogeneous parts in (\ref{eq:eom_det}-\ref{eq:eom_stoc}). Clearly, $\boldsymbol{\Omega}_{\rm D}$ and $\boldsymbol{\Omega}_{\rm S}$ are the deterministic and stochastic parts of $\boldsymbol{\Omega}$, respectively. The deterministic part $\boldsymbol{\Omega}_{\rm D}$ can be dealt with using methods for solving linear ordinary differential equations. We have
\begin{equation}
\label{eq:mean_OmegaD}
    \boldsymbol{\Omega}_{\rm D}(t) = e^{-\mathbf{D}_\eta t} \boldsymbol{\Omega}_0 + \int_0^t dt_1 \, e^{-\mathbf{D}_\eta (t-t_1)} \mathbf{N} \,,
\end{equation}
where $\boldsymbol{\Omega}_0$ is an initial prior.

The stochastic part $\boldsymbol{\Omega}_{\rm S}$ can be recognized as a bivariate OU process with a drift matrix $\mathbf{D}_\eta$ and a diffusion matrix $\mathbf{Q}$. 
The probability $P(\boldsymbol{\Omega}_{\rm S}, t | \boldsymbol{\Omega}_{\rm S}', t')$ of observing $\boldsymbol{\Omega}_{\rm S}$ at time $t$ and $\boldsymbol{\Omega}_{\rm S}'$ at time $t'$ satisfies the Fokker-Planck equation:
\begin{equation}
\label{eq:fp_2cm}
    \partial_t P(\boldsymbol{\Omega}_{\rm S}, t | \boldsymbol{\Omega}_{\rm S}', t') = {\cal L}[P(\boldsymbol{\Omega}_{\rm S}, t | \boldsymbol{\Omega}_{\rm S}', t')] \,,
\end{equation}
where the functional ${\cal L}[F(\boldsymbol{\Omega}_{\rm S})]$ is given by
\begin{equation}
\label{eq:fp_func_2cm}
    {\cal L}[F(\boldsymbol{\Omega}_{\rm S})] \equiv \dfrac{\partial}{\partial \Omega_{{\rm S},i}} \left( D_{ij} \Omega_{{\rm S},j} F(\boldsymbol{\Omega}_{\rm S}) \right) + \dfrac{1}{2} \dfrac{\partial^2}{\partial \Omega_{{\rm S},i} \partial \Omega_{{\rm S},j}} \left( Q_{ij} F(\boldsymbol{\Omega}_{\rm S}) \right) \,.
\end{equation}
The solution is a Gaussian distribution $P(\boldsymbol{\Omega}_{\rm S}, t | \boldsymbol{\Omega}_{\rm S}', t') \sim {\cal N}\left( \boldsymbol{\mu}, \boldsymbol{\Sigma} \right)$ with mean $\boldsymbol{\mu}$ and covariance $\boldsymbol{\Sigma}$ given by
\begin{align}
    \label{eq:lambda_2cm_def}
    \boldsymbol{\Lambda} &= e^{-\mathbf{D}_\eta |t - t'|} \\
    \label{eq:mean_2cm_stoc}
    \boldsymbol{\mu} &= \boldsymbol{\Lambda} \boldsymbol{\Omega}_{\rm S}' \\
    \label{eq:cov_2cm_stoc}
    \boldsymbol{\Sigma} &= \mathbf{c} - \boldsymbol{\Lambda} \mathbf{c} \boldsymbol{\Lambda}^\top \,.
\end{align}
This is valid for arbitrary times $t$ and $t'$. The matrix $\mathbf{c}$ is the solution to the Lyapunov equation:
\begin{equation}
\label{eq:lyapunov_2cm}
    \mathbf{D}_\eta \mathbf{c} + \mathbf{c} \mathbf{D}_\eta^\top = \mathbf{Q} \,.
\end{equation}
The Lyapunov equation can be solved analytically for the regularized drift matrix $\mathbf{D}_\eta$. The stationary solution is given by
\begin{equation}
    \boldsymbol{\Omega}_{\rm S} \sim {\cal N}\left( \boldsymbol{0}, \mathbf{c} \right) \,.
\end{equation}
This solution anchors a fast Bayesian inference method to be introduced in the following section \citep{2018PhRvE..98a2136S}.

The covariance of the stochastic part $\boldsymbol{\Omega}_{\rm S}$ determines the covariance of the process $\boldsymbol{\Omega}$; since the deterministic and stochastic parts are uncorrelated. We have
\begin{align}
    \langle \boldsymbol{\Omega} \rangle = \boldsymbol{\Omega}_{\rm D} + \langle \boldsymbol{\Omega}_{\rm S} \rangle 
\end{align}
and
\begin{align}
    \langle \boldsymbol{\Omega} \boldsymbol{\Omega}^\top \rangle = \boldsymbol{\Omega}_{\rm D} \boldsymbol{\Omega}_{\rm D}^\top + \boldsymbol{\Omega}_{\rm D} \langle  \boldsymbol{\Omega}_{\rm S}^\top \rangle + \langle \boldsymbol{\Omega}_{\rm S} \rangle \boldsymbol{\Omega}_{\rm D}^\top + \langle \boldsymbol{\Omega}_{\rm S} \boldsymbol{\Omega}_{\rm S}^\top \rangle \,.
\end{align}
At long times compared to the initial time, $t - t_0 \gg \tau$, the mean of the stochastic contribution vanishes, $\langle \boldsymbol{\Omega}_{\rm S} \rangle \rightarrow 0$. Then, the mean and the covariance can be written as
\begin{align}
    \langle \boldsymbol{\Omega} \rangle \sim \boldsymbol{\Omega}_{\rm D}
\end{align}
and
\begin{align}
    \langle \boldsymbol{\Omega} \boldsymbol{\Omega}^\top \rangle \sim \boldsymbol{\Omega}_{\rm D} \boldsymbol{\Omega}_{\rm D}^\top + \langle \boldsymbol{\Omega}_{\rm S} \boldsymbol{\Omega}_{\rm S}^\top \rangle \,,
\end{align}
with $\langle \boldsymbol{\Omega}_{\rm S} \boldsymbol{\Omega}_{\rm S}^\top \rangle \sim \mathbf{c}$ where $\mathbf{c}$ is the solution of the Lyapunov equation \eqref{eq:lyapunov_2cm}.
This gives a regularized solution to (\ref{eq:eom_crust}-\ref{eq:xic_xis}) that is rooted in established theory \citep{2018PhRvE..98a2136S}. In addition, time-domain analytical (exact and asymptotic) solutions to the frequencies and phases have been expressed in \cite{Bernardo:2026opv}. These will be utilized in Section \ref{sec:fast_bayesian} to develop scalable Gaussian process models of spin noise.

\subsection*{The stationary limit}

The four-parameter process (\ref{eq:Jeom_minimal}-\ref{eq:wncov_minimal}) characterized by a conserved total angular momentum \citep{Antonelli:2022gqw} can be considered as the stationary limit of the six-parameter process (\ref{eq:eom_crust}-\ref{eq:xic_xis}). A regularized solution can be teased out analytically following the method outlined previously. In vector form \eqref{eq:eom_2cm_matrix}, the stationary model (\ref{eq:Jeom_minimal}-\ref{eq:wncov_minimal}) is given by the drift, torque, and volatility matrices
\begin{equation}
\label{eq:drift_matrix_2cm_minimal}
    \mathbf{D} = \begin{bmatrix} x_{\rm s}/\tau & -x_{\rm s}/\tau \\ -x_{\rm c}/\tau & x_{\rm c}/\tau \end{bmatrix} \,,
\end{equation}
\begin{equation}
\label{eq:torque_matrix_2cm_minimal}
    \mathbf{N} = \begin{bmatrix} N_{\rm c}/I_{\rm c} \\ 0 \end{bmatrix} \,,
\end{equation}
\begin{equation}
\label{eq:sigma_matrix_minimal}
    \boldsymbol{\sigma} = \sqrt{Q}\begin{bmatrix} 1/x_{\rm c} & 0 \\ -1/x_{\rm s} & 0 \end{bmatrix} \,.
\end{equation}

However, the solution to (\ref{eq:Jeom_minimal}-\ref{eq:wncov_minimal}) can also be expressed simply. At large times compared to the relaxation time, the stationary, Gaussian process can be fully determined by the angular momentum and the first two moments of the crust-core lag:
\begin{align}
    J &\sim N_{\rm c} t \\
    \langle \Omega_-(t) \rangle &\sim \dfrac{\tau N_{\rm c}}{I_{\rm c}} \\
    \langle \Omega_-(t) \Omega_-(t') \rangle - \langle \Omega_-(t) \rangle \langle \Omega_-(t') \rangle &\sim  \dfrac{Q \tau}{2 (x_{\rm c} x_{\rm s})^2} e^{-|t' - t| / \tau} \,.
\end{align}
This can be translated to component form.\footnote{In terms of the total angular momenta and the crust-core lag, the crust and superfluid angular velocities can be written as $\Omega_{\rm c} = x_{\rm s} \Omega_- + (J/I)$ and $\Omega_{\rm s} = -x_{\rm c} \Omega_- + (J/I)$, respectively.} For the crust, we have
\begin{align}
    \langle \Omega_{\rm c}(t) \rangle &\sim \dfrac{\tau x_{\rm s} N_{\rm c}}{I_{\rm c}} + \dfrac{N_{\rm c}}{I} t \\
    \langle \Omega_{\rm c}(t) \Omega_{\rm c}(t') \rangle - \langle \Omega_{\rm c}(t) \rangle \langle \Omega_{\rm c}(t') \rangle &\sim \dfrac{Q \tau}{2 x_{\rm c}^2} e^{-|t' - t| / \tau} \,,
\end{align}
and for the superfluid, we have
\begin{align}
    \langle \Omega_{\rm s}(t) \rangle &\sim -\dfrac{\tau x_{\rm c} N_{\rm c}}{I_{\rm c}} + \dfrac{N_{\rm c}}{I} t \\
    \langle \Omega_{\rm s}(t) \Omega_{\rm s}(t') \rangle - \langle \Omega_{\rm s}(t) \rangle \langle \Omega_{\rm s}(t') \rangle &\sim \dfrac{Q \tau}{2 x_{\rm s}^2} e^{-|t' - t| / \tau} \,.
\end{align}

The above solutions make desirable features of (\ref{eq:Jeom_minimal}-\ref{eq:wncov_minimal}) transparent. First, in contrast with (\ref{eq:eom_crust}-\ref{eq:xic_xis}), the angular velocity residuals are a stationary process. The spin-down of a pulsar can be entirely associated with the constant external torque $N_{\rm c}$, and spin wandering to mutual friction and the stochastic torque $\xi(t)$. The component residuals are described by an OU process with the same relaxation time and amplitudes differentiated only by each components' moment of inertia. It can be shown that $\delta \Omega_{\rm c}(t) \sim - (x_{\rm s}/x_{\rm c}) \delta \Omega_{\rm s} (t)$ where $\delta \Omega_i(t) = \Omega_i(t) - \langle \Omega_i(t) \rangle$ are corresponding zero mean processes. The negative sign implies that the frequency residuals of the crust and the superfluid are anticorrelated. Note that if $x_{\rm s} \ll 1$ and $x_{\rm c} \sim 1$, the superfluid angular velocity residuals will at least two orders of magnitude above the corresponding crust velocity residuals \citep{2012PhRvL.109x1103A}. This sets down a path to falsifying the stationary model.
Radio timing noise tied to the crust rotational state will be sensitive to spin wandering through the two-parameter combination $(\tau, A_{\rm min} = Q \tau / x_{\rm c}^2)$.
Glitches constrain the fractional moment of inertia $x_{\rm s}$ \citep{2020A&A...642A.223M}.
Then, an independent observation of the superfluid angular velocity residuals provides a consistency check of, or falsify, the stationary process (\ref{eq:Jeom_minimal}-\ref{eq:wncov_minimal}) as a physical model of spin noise.

\section{Fast Bayesian inference methods}
\label{sec:fast_bayesian}

The regularized two-component model (\ref{eq:eom_2cm_matrix}) with drift matrix \eqref{eq:drift_matrix_2cm_reg} constitutes a bivariate OU process with strictly positive eigenvalues. This enables application of the fast Bayesian inference method in \cite{2018PhRvE..98a2136S} for multivariate OU processes in the idealized scenario that both $\Omega_{\rm c}$ and $\Omega_{\rm s}$ are observed and that the data is sampled evenly. The sufficient statistics must be computed from the stochastic residuals $\boldsymbol{\Omega}_{\rm S} = \boldsymbol{\Omega} - \boldsymbol{\Omega}_{\rm D} = [\delta \Omega_{\rm c}, \delta \Omega_{\rm s}]^\top$; where $\delta \Omega_{\rm c}$ and $\delta \Omega_{\rm s}$ are the stochastic spin wandering components of the crust and superfluid angular velocities, respectively. When only the crust angular velocity is observed, we rely on analytic time-domain solution derived in \cite{Bernardo:2026opv} to construct a scalable Gaussian process method.

\subsection{Sufficient statistics}
\label{subsec:sufficient_statistics}

Consider discrete observations of the stochastic residuals $\boldsymbol{\Omega}_{{\rm S},n} \equiv \boldsymbol{\Omega}_n - \boldsymbol{\Omega}_{{\rm D},n}$ at times $t_n = n\Delta t$ ($n=1,\ldots,N$):
$
\boldsymbol{X} = \left\{ \boldsymbol{\Omega}_{{\rm S},1}, \boldsymbol{\Omega}_{{\rm S},2}, \ldots, \boldsymbol{\Omega}_{{\rm S},N} \right\},
$
where $\boldsymbol{\Omega}_{{\rm S},n} = [\delta\Omega_{{\rm c}}(t_n), \delta \Omega_{{\rm s}}(t_n)]^\top$. Utilizing the Markov property of the OU process, the likelihood of the observed stochastic residuals given parameters $\boldsymbol{\theta} = (\tau_{\rm c}, \tau_{\rm s}, Q_{\rm c}, Q_{\rm s})$ can be factorized as
\begin{equation}
\label{eq:likelihood_markov}
    P(\boldsymbol{X}|\boldsymbol{\theta}) = P(\boldsymbol{\Omega}_{{\rm S},1}|\boldsymbol{\theta}) \prod_{n=1}^{N-1} P(\boldsymbol{\Omega}_{{\rm S},n+1} | \boldsymbol{\Omega}_{{\rm S},n}, \boldsymbol{\theta}) \,.
\end{equation}
Bayes theorem gives the posterior distribution over parameters $\boldsymbol{\theta}$ as
\begin{equation}
    P(\boldsymbol{\theta}|\boldsymbol{X}) = \frac{P(\boldsymbol{X}|\boldsymbol{\theta}) P(\boldsymbol{\theta})}{P(\boldsymbol{X})} \,.
\end{equation}
Considering uniform priors over the parameters, the posterior, proportional to the likelihood, $P(\boldsymbol{\theta}|\boldsymbol{X}) \propto P(\boldsymbol{X}|\boldsymbol{\theta})$, can be evaluated analytically as (\cite{2018PhRvE..98a2136S})
\begin{equation}
\begin{aligned}
& \ln P(\boldsymbol{\theta}\mid \boldsymbol{X}) \\
& = -\frac{1}{2}\operatorname{tr}\Bigg\{
\boldsymbol{\Sigma}^{-1}\Big[
\big(\boldsymbol{\Lambda}-\mathbf{T}_{2}\mathbf{T}_{3}^{-1}\big)\mathbf{T}_{3}
\big(\boldsymbol{\Lambda}-\mathbf{T}_{2}\mathbf{T}_{3}^{-1}\big)^{\top}
+\big(\mathbf{T}_{1}-\mathbf{T}_{2}\mathbf{T}_{3}^{-1}\mathbf{T}_{2}^{\top}\big)
\Big]\Bigg\} \\
&\quad -\frac{N-1}{2}\ln\!\left|2\pi\boldsymbol{\Sigma}\right|
-\frac{1}{2}\operatorname{tr}\!\left(\mathbf{c}^{-1}\mathbf{T}_{4}\right)
-\frac{1}{2}\ln\!\left|2\pi\mathbf{c}\right| \,,
\end{aligned}
\label{eq:postBayesI}
\end{equation}
where the four $2\times2$ sufficient statistic matrices are given by
\begin{align}
    \mathbf{T}_1 &= \sum_{n=1}^{N-1} \boldsymbol{\Omega}_{{\rm S},n+1} \boldsymbol{\Omega}_{{\rm S},n+1}^\top \,, \label{eq:T1} \\
    \mathbf{T}_2 &= \sum_{n=1}^{N-1} \boldsymbol{\Omega}_{{\rm S},n+1} \boldsymbol{\Omega}_{{\rm S},n}^\top \,, \label{eq:T2} \\
    \mathbf{T}_3 &= \sum_{n=1}^{N-1} \boldsymbol{\Omega}_{{\rm S},n} \boldsymbol{\Omega}_{{\rm S},n}^\top \,, \label{eq:T3} \\
    \mathbf{T}_4 &= \boldsymbol{\Omega}_{{\rm S},1} \boldsymbol{\Omega}_{{\rm S},1}^\top \,. \label{eq:T4}
\end{align}
These matrices encode all information needed for parameter estimation and can be computed in ${\cal O}(N)$ time.

\subsubsection*{Maximum a posteriori (MAP) estimation}

The MAP estimates for the transition matrix $\boldsymbol{\Lambda} = e^{-\mathbf{D}_\eta \Delta t}$ and the covariance matrix $\boldsymbol{\Sigma}$ are:
\begin{align}
    \boldsymbol{\Lambda}^\ast &= \mathbf{T}_2 \mathbf{T}_3^{-1} \,, \label{eq:Lambda_MAP} \\
    \boldsymbol{\Sigma}^\ast &= \frac{1}{N} \left( \mathbf{T}_1 - \mathbf{T}_2 \mathbf{T}_3^{-1} \mathbf{T}_2^\top \right) \,. \label{eq:Sigma_MAP}
\end{align}
The drift matrix $\mathbf{D}_\eta^\ast$ is recovered via the matrix logarithm,
$
\mathbf{D}_\eta^\ast = -\ln \boldsymbol{\Lambda}^\ast / \Delta t \,,
$
and the stationary covariance $\mathbf{c}^\ast$ is obtained by numerically solving the Lyapunov Stein equation \eqref{eq:cov_2cm_stoc} at the MAP $\boldsymbol{\Lambda}^*$ and $\boldsymbol{\Sigma}^*$.

The physical parameters are then extracted as follows. The diagonal entries of the regularized drift matrix \eqref{eq:drift_matrix_2cm_reg} are exactly $D_{\eta,11} = 1/\tau_{\rm c}$ and $D_{\eta,22} = 1/\tau_{\rm s}$ for any value of $\eta$, so the individual relaxation times are read off directly from the MAP estimate:
\begin{align}
    \tau_{\rm c}^\ast = 1/D_{11}^\ast \,, \quad \tau_{\rm s}^\ast = 1/D_{22}^\ast \,,
\end{align}
from which $\tau^\ast$ and $r^\ast$ follow via (\ref{eq:tau_def}--\ref{eq:r_def}):
\begin{align}
    \tau^\ast &= \frac{\tau_{\rm c}^\ast \tau_{\rm s}^\ast}{\tau_{\rm c}^\ast + \tau_{\rm s}^\ast} \,, \quad r^\ast = \frac{\tau_{\rm s}^\ast}{\tau_{\rm c}^\ast} \,.
\end{align}
The diffusion coefficients follow from the diagonal elements of the Lyapunov equation \eqref{eq:lyapunov_2cm} evaluated at $\mathbf{D}_\eta^\ast$ and $\mathbf{c}^\ast$:
\begin{align}
    Q_{\rm c}^\ast &= \frac{2 c_{11}^\ast}{\tau_{\rm c}^\ast} + 2c_{12}^\ast \left( \frac{1}{\eta} - \frac{1}{\tau_{\rm c}^\ast} \right), \label{eq:Qc_MAP} \\
    Q_{\rm s}^\ast &= \frac{2 c_{22}^\ast}{\tau_{\rm s}^\ast} + 2c_{12}^\ast \left( \frac{1}{\eta} - \frac{1}{\tau_{\rm s}^\ast} \right). \label{eq:Qs_MAP}
\end{align}
The off-diagonal element of the Lyapunov equation provides a self-consistency check on $\mathbf{c}^\ast$:
\begin{equation}
    c_{12}^\ast \left( \frac{1}{\tau_{\rm c}^\ast} + \frac{1}{\tau_{\rm s}^\ast} \right) + \frac{c_{11}^\ast + c_{22}^\ast}{\eta} - \frac{c_{22}^\ast}{\tau_{\rm c}^\ast} - \frac{c_{11}^\ast}{\tau_{\rm s}^\ast} = 0. \label{eq:lyapunov_check}
\end{equation}
The steady-state quantities $\llangle \alpha_{\rm c} \rrangle$ and $\llangle \Delta \Omega \rrangle$ are recovered by fitting the deterministic solution \eqref{eq:mean_OmegaD} to the sample mean $\overline{\boldsymbol{\Omega}}$, which yields $N_{\rm c}/I_{\rm c}$ and $N_{\rm s}/I_{\rm s}$, and hence $\llangle \alpha_{\rm c} \rrangle$ and $\llangle \Delta \Omega \rrangle$ through (\ref{eq:alpha_ss}--\ref{eq:lag_ss}).

\subsubsection*{Model comparison}

We have assumed that the regularized two-component model is
given and that our task is to estimate its parameters. In practice, the model itself may be uncertain, and one may wish to compare results with different values of the regularization parameter
$\eta$ or compare the two-component model with alternatives; such as to test stationarity.
The probability of a model ${\cal M}_\alpha$ given the data $\boldsymbol{X}$ is
\begin{equation}
    P(\mathcal{M}_\alpha | \boldsymbol{X}) \propto P(\boldsymbol{X} | \mathcal{M}_\alpha)\,P(\mathcal{M}_\alpha) \,,
\end{equation}
where the first factor is the evidence for the model and the second
its prior probability. Assuming all models to be a priori equally likely, the
evidence is the normalizing constant obtained by marginalizing the likelihood
over the parameter space $\boldsymbol{\theta}$:
\begin{equation}
    P(\boldsymbol{X} | \mathcal{M}_\alpha) = \int d\boldsymbol{\theta}\, P(\boldsymbol{X} |
    \boldsymbol{\theta}, \mathcal{M}_\alpha)\,P(\boldsymbol{\theta} |
    \mathcal{M}_\alpha) \,.
\end{equation}
For unimodal posteriors, the height at the MAP value $\boldsymbol{\theta}^\ast$
times the width $\Delta\boldsymbol{\theta}$ provides a good approximation for
the evidence \citep{2018PhRvE..98a2136S}:
\begin{equation}
\label{eq:evidence}
    P(\boldsymbol{X} | \mathcal{M}_\alpha) \simeq
    P(\boldsymbol{X} | \boldsymbol{\theta}^\ast, \mathcal{M}_\alpha)\,
    P(\boldsymbol{\theta}^\ast | \mathcal{M}_\alpha)\,\Delta\boldsymbol{\theta} \,.
\end{equation}
The first factor is the best-fit likelihood. The second or the product of
the prior for the MAP estimate and its standard error is called the
Ockham factor.

The width $\Delta\boldsymbol{\theta}$ is determined by the Hessian matrix
$\mathbf{A} = -\nabla\nabla\ln P(\boldsymbol{\theta}|\boldsymbol{X})$
evaluated at the MAP. It can be shown that \citep{2003itil.book.....M}:
\begin{equation}
\label{eq:evidence_laplace}
    \ln P(\boldsymbol{X}|\mathcal{M}_\alpha) \simeq
    \ln P(\boldsymbol{X}|\boldsymbol{\theta}^\ast) +
    \ln P(\boldsymbol{\theta}^\ast) -
    \dfrac{1}{2}\ln\,|\mathbf{A}/(2\pi)| \,.
\end{equation}
The Hessian $\mathbf{A}$, partitioned into blocks over
$(\boldsymbol{\Sigma}^{-1}, \boldsymbol{\Lambda})$ at the MAP estimates
(\ref{eq:Lambda_MAP}--\ref{eq:Sigma_MAP}), is \citep{2018PhRvE..98a2136S}:
\begin{align}
    \mathbf{A}_{11} &= \frac{N}{2}(\boldsymbol{\Sigma}^\ast)^2 \,, \\
    \mathbf{A}_{12} &= \frac{N}{2}
        \bigl(\boldsymbol{\Lambda}^\ast\mathbf{c}^\ast +
        \mathbf{c}^\ast\boldsymbol{\Lambda}^{\ast\top}\bigr) \,, \\
    \mathbf{A}_{22} &= \boldsymbol{\Sigma}^{\ast-1}
        \bigl(\mathbf{T}_3 - N\mathbf{c}^\ast\bigr) -
        \frac{N}{2}\boldsymbol{\Sigma}^{\ast-2}
        \bigl(\boldsymbol{\Lambda}^\ast\mathbf{c}^\ast +
        \mathbf{c}^\ast\boldsymbol{\Lambda}^{\ast\top}\bigr)^2 \,.
\end{align}
The logarithm of the evidence evaluates to
\citep{2018PhRvE..98a2136S, 2003itil.book.....M}:
\begin{equation}
\label{eq:log_evidence_explicit}
\begin{aligned}
    \ln P(\boldsymbol{X}|\mathcal{M}_\alpha) \simeq{}&
    -\frac{N}{2}
        \ln \left((2\pi e)^M|\boldsymbol{\Sigma}^\ast|\right) \\
    &-\dfrac{1}{2}\ln \left(
        |\mathbf{A}_{22}|\, 
        |\mathbf{A}_{11} - \mathbf{A}_{12}\mathbf{A}_{22}^{-1}
        \mathbf{A}_{21}|
    \right) + \dfrac{M}{2} \ln 2\pi \,,
\end{aligned}
\end{equation}
where $M=2$ for the two-component model.

The Laplace approximation (\ref{eq:evidence}-\ref{eq:log_evidence_explicit}) assumes a unimodal posterior that is well approximated by a Gaussian around $\boldsymbol\theta^\ast$, with the MAP estimate lying away from any prior boundary; it can be inaccurate for multimodal or boundary-peaked posteriors, which should be checked against the full posterior samples used elsewhere in this work. It is also worth clarifying the scalability of the different pieces of this framework: the sufficient statistics (\ref{eq:T1}-\ref{eq:T4}) and the MAP/Laplace evidence above are ${\cal O}(N)$ for the full-state (crust and superfluid) likelihood, whereas the analytic semiseparable likelihood of Section~\ref{subsec:analytic_semiseparable} extends this ${\cal O}(N)$ scaling to the crust-only case; the ${\cal O}(N^3)$ direct-GP evidence reported in Sections~\ref{sec:singular_simulations}-\ref{sec:minimal_simulations} is used only as a full-state cross-check, not as the scalable endpoint of the method.

Appendix \ref{sec:2d_brownian} sets a baseline example of the above method applied to 2D Brownian motion.

\begin{figure*}
    \centering
    \includegraphics[width=0.94\textwidth]
        {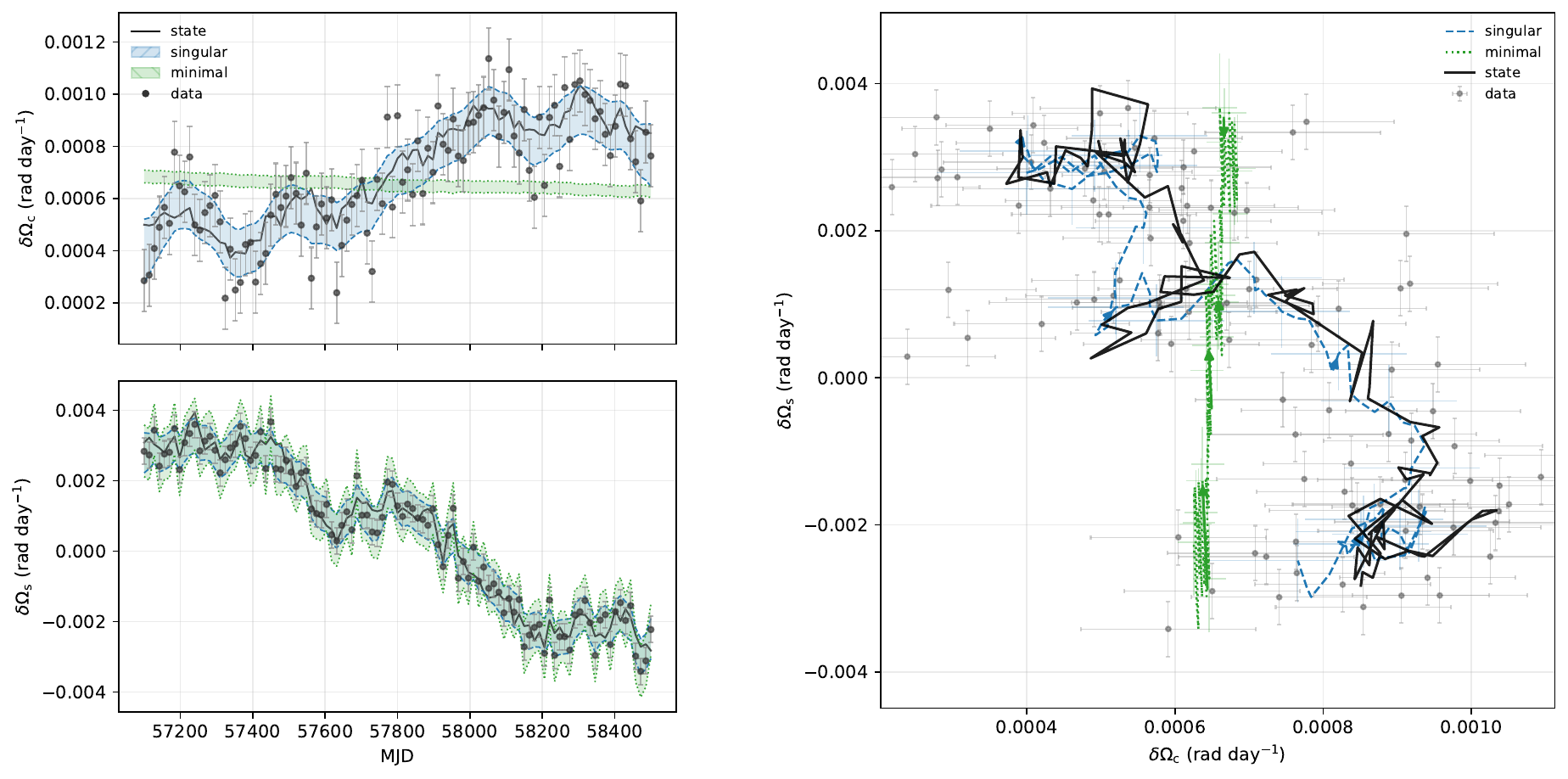}
    \includegraphics[width=0.94\textwidth]
        {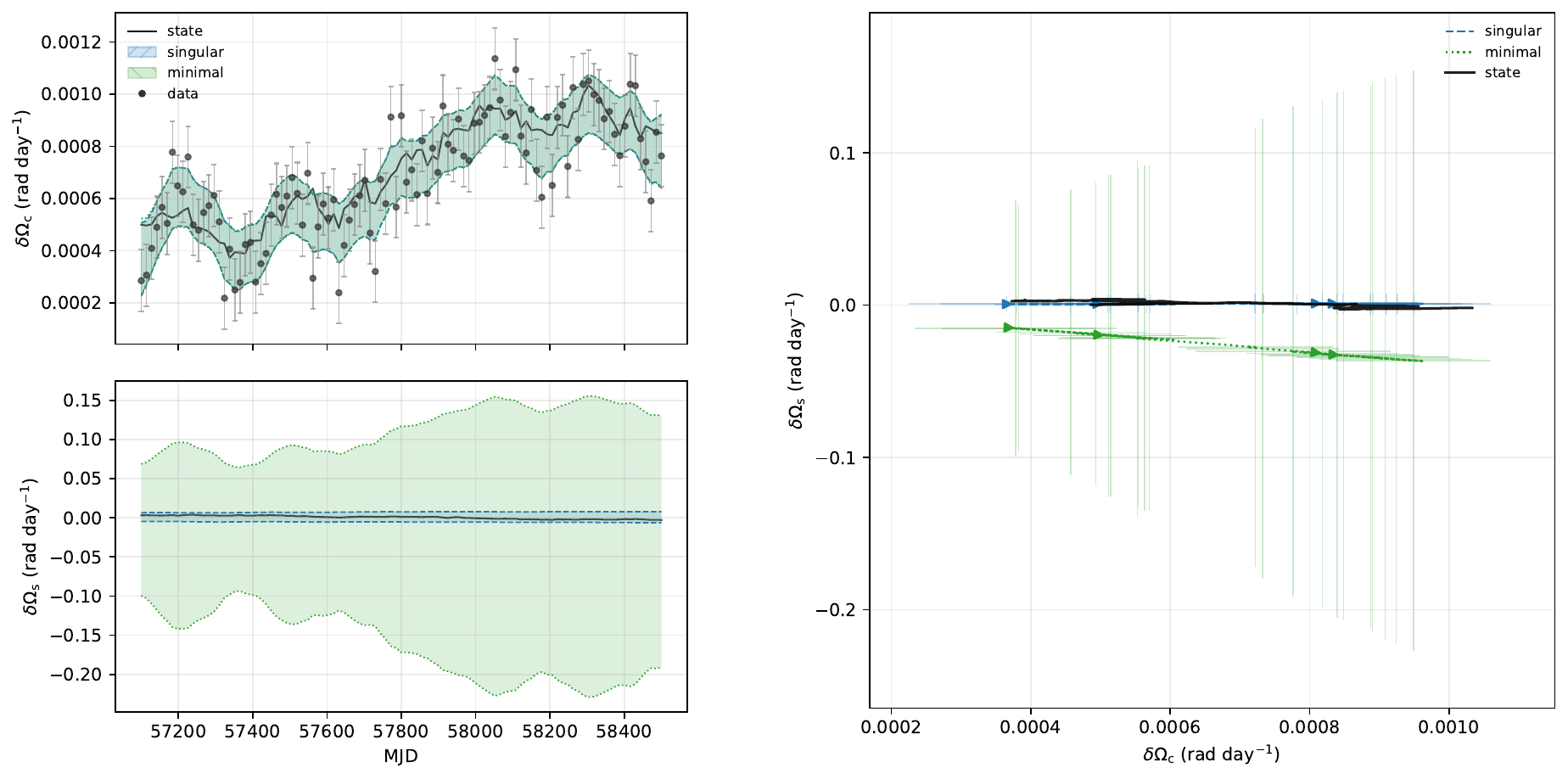}
    \caption{
    Reconstruction of a singular model realization using the singular
    and minimal models. The top figure uses full-state observations of
    the crust and superfluid. The lower figure uses crust-only
    observations. The left panels show the component time series and
    95 percent reconstruction bands, while the right panels show the
    2D trajectories. The black curve is
    the true state, the points with error bars are the mock data, and the
    blue and green curves denote the singular and minimal reconstructions,
    respectively.}
    \label{fig:singular_reconstruction}
\end{figure*}

\begin{figure*}
    \centering
    \includegraphics[width=0.56\textwidth]
        {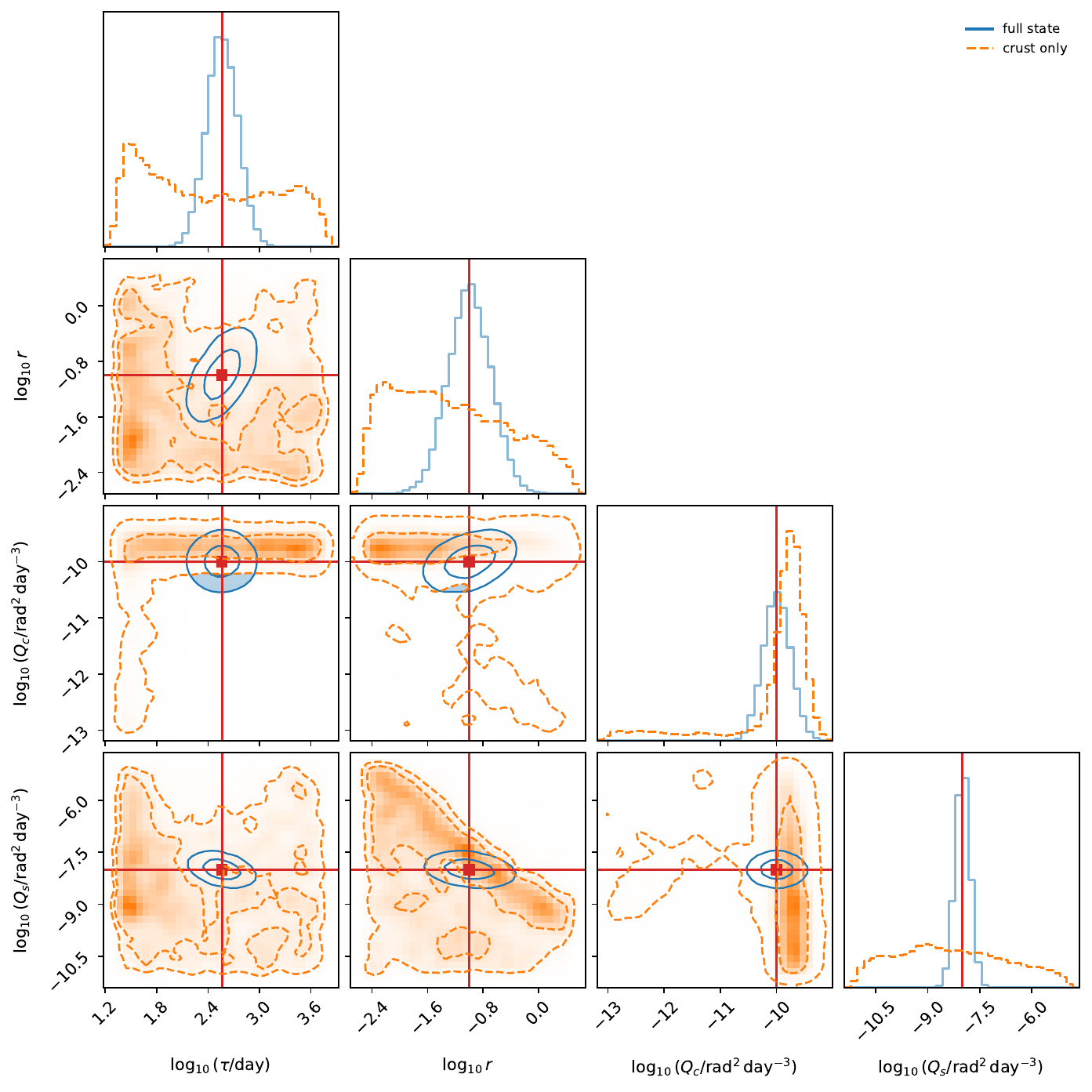}
    \hfill
    \includegraphics[width=0.41\textwidth]
        {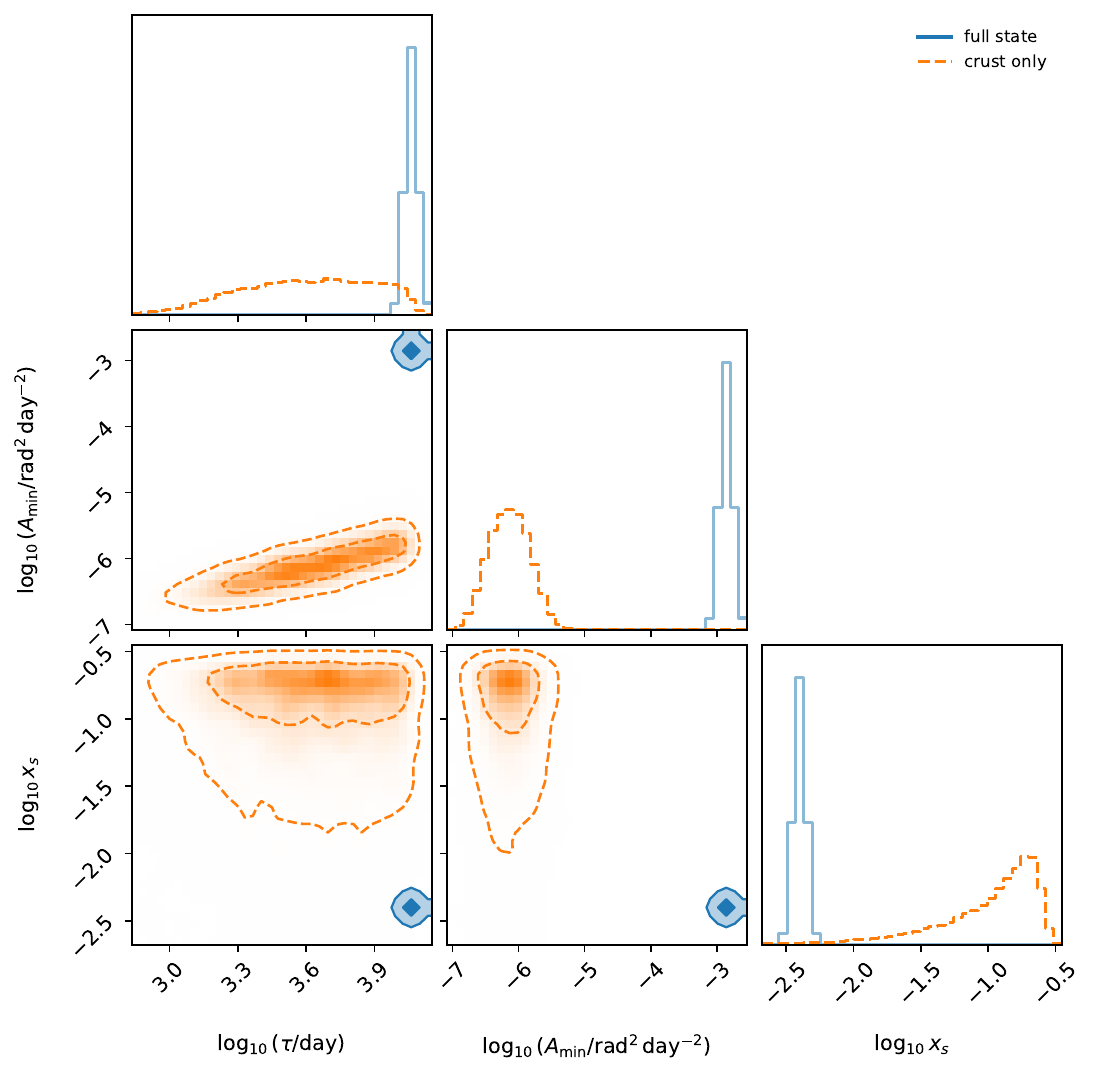}
    \caption{
    Parameter posteriors for the singular model realization (Table \ref{tab:singular_parameters}) analyzed
    with the singular model (left) and the minimal model (right).
    Blue solid distributions use full-state observations and orange
    dashed distributions use crust-only observations. Red lines mark
    the injected parameters only in the generating-model panel. Contours
    enclose 68 and 95 percent posterior probability.}
    \label{fig:singular_posteriors}
\end{figure*}

\begin{figure*}
    \centering
    \includegraphics[width=0.88\textwidth]
        {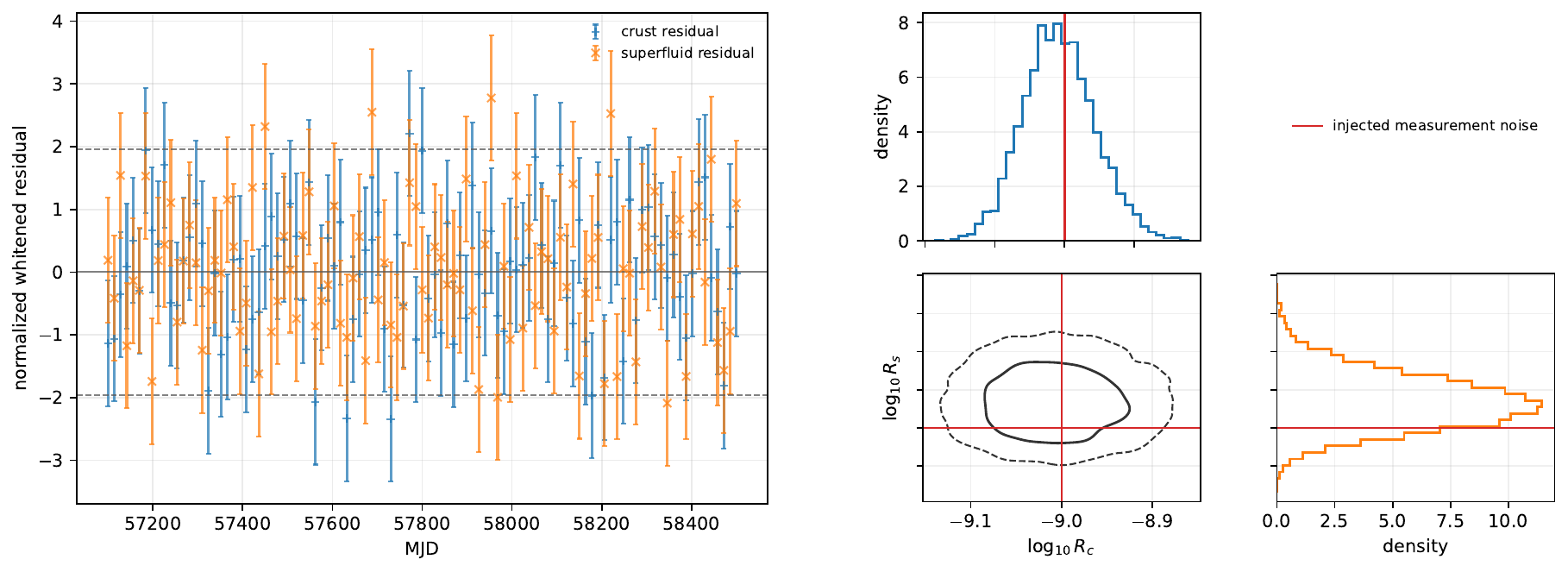}
    \includegraphics[width=0.88\textwidth]
        {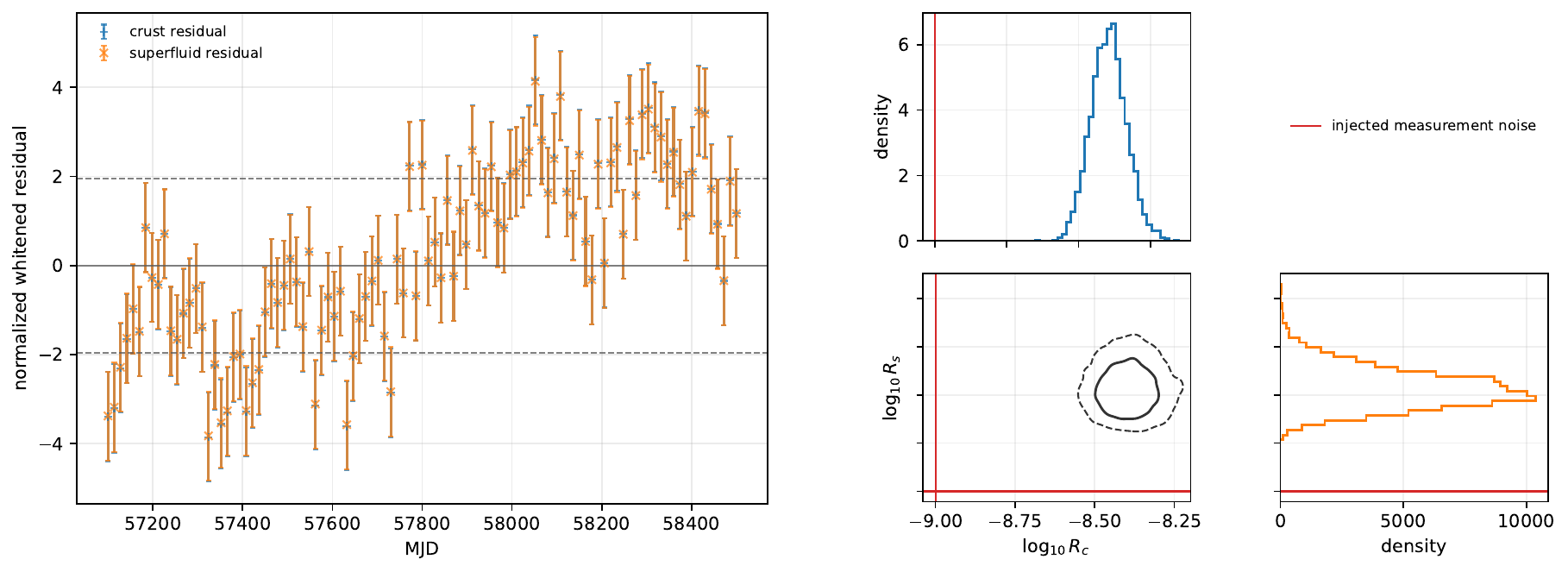}
    \includegraphics[width=0.88\textwidth]
        {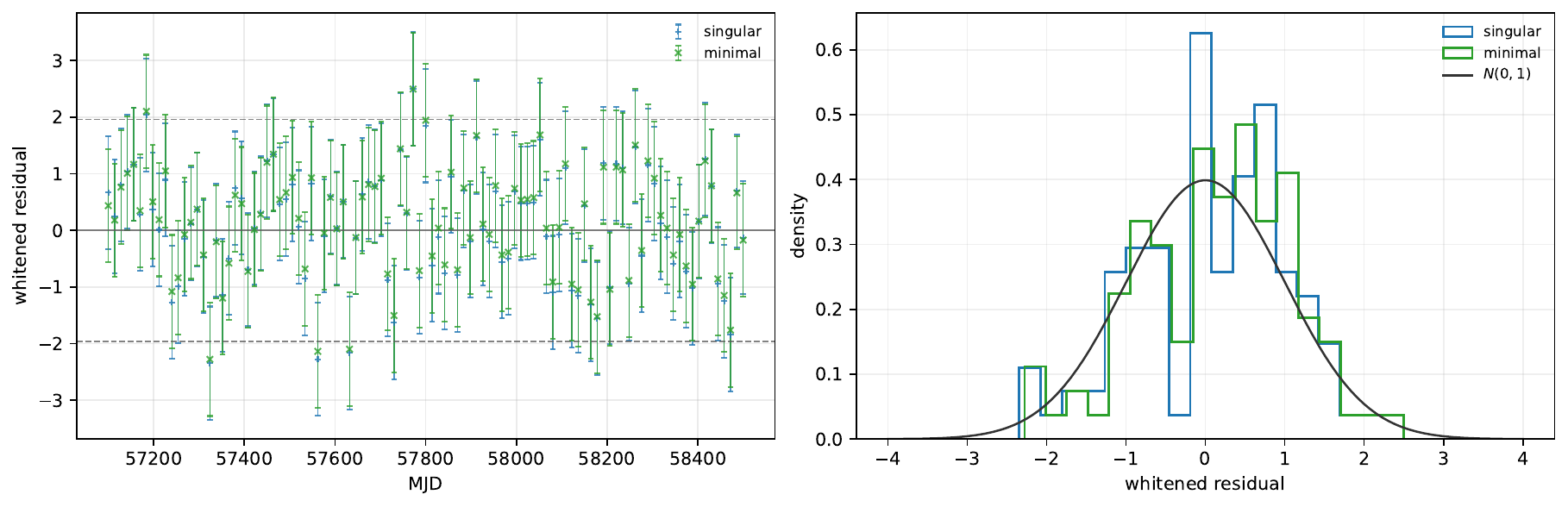}
    \caption{
    Noise diagnostics for the singular model realization. From top to
    bottom, the figures show the full-state singular fit, the full-state
    minimal fit, and the crust-only whitened residuals. In the full-state
    panels, $+$ and $\times$ mark the normalized crust and superfluid
    residuals. The right panels show the residual-based estimates of
    $R_{\rm c}$ and $R_{\rm s}$, with solid and dashed contours enclosing
    68 and 95 percent probability. Red lines mark the
    measurement noise parameters.}
    \label{fig:singular_noise}
\end{figure*}

\begin{figure*}
    \centering
    \includegraphics[width=0.94\textwidth]
        {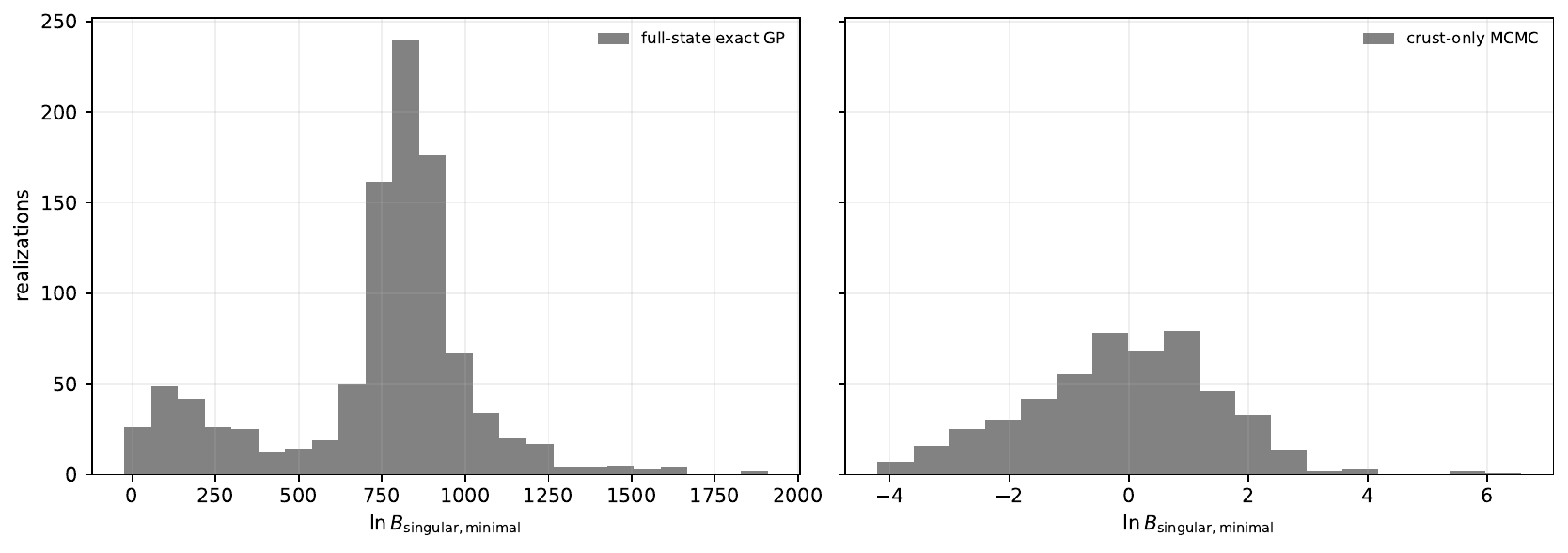}
    \caption{
    Distribution of $\ln B_{\rm singular,minimal}$ for data generated
    with the singular model. The left panel contains 1000 full-state
    exact-GP results and the right panel contains 500 crust-only
    MCMC stepping-stone sampling results. Positive values favour the
    singular model.}
    \label{fig:singular_bayes_factors}
\end{figure*}

\subsection{Analytic semiseparable likelihood}
\label{subsec:analytic_semiseparable}

The fast Bayesian method in the previous sections assumes evenly sampled observations of the full stochastic state
$\boldsymbol{\Omega}_{\rm S}=(\delta\Omega_{\rm c},\delta\Omega_{\rm s})^\top$.
Real radio timing data do not satisfy these assumptions.  The observations are irregularly sampled, heteroscedastic, and directly probe only the crustal angular velocity.
We construct an independent fast likelihood from the analytical covariance of the crust alone, using the time-domain solutions derived in \cite{Bernardo:2026opv}.  The resulting method is analogous in spirit to
\cite{2017AJ....154..220F}; instead of forming and factorizing a dense covariance matrix, we utilize a semiseparable representation that permits the action of the inverse covariance and the logarithm of its determinant to be evaluated in ${\cal O}(N)$ operations. The full-state algorithm is given in Appendix \ref{sec:full_state_semiseparable}.

Let $t_i$ denote radio-frequency epochs and let
\begin{equation}
    y_i = \delta\Omega_{\rm c}(t_i) + \sum_{\mu=0}^{p-1} M_{i\mu}\beta_\mu
    + \varepsilon_i
\end{equation}
be the observed crust angular velocity residual.  Here $M_{i\mu}$ is a low-order deterministic trend design matrix, e.g. $M_{i\mu}=(1,t_i-t_{\rm ref})$ for a constant and linear spin-frequency trend, $\beta_\mu$ are nuisance trend parameters, and
\begin{equation}
    \langle \varepsilon_i\varepsilon_j\rangle = \sigma_i^2\delta_{ij}
\end{equation}
is the heteroscedastic radio measurement covariance.  The stochastic covariance entering the likelihood is therefore
\begin{equation}
    C_{ij}(\boldsymbol{\theta}) =
    K_{\rm c}(t_i,t_j;\boldsymbol{\theta}) + \sigma_i^2\delta_{ij} \,,
\end{equation}
where $\boldsymbol{\theta}=(\tau,r,Q_{\rm c},Q_{\rm s})$ and
$K_{\rm c}$ is the finite-start analytical covariance of the crustal angular velocity.

The useful structure appears most transparently in the plus and minus modes. Define
\begin{equation}
    \tau\Omega_+ = \tau_{\rm c}\Omega_{\rm c}+\tau_{\rm s}\Omega_{\rm s} \,,
    \qquad
    \Omega_- = \Omega_{\rm c}-\Omega_{\rm s} \,.
\end{equation}
The mode $\Omega_-$ is the damped crust-core lag and $\Omega_+$ is the diffusive total angular momentum mode.  Writing $x=t-t_0$ and $x'=t'-t_0$, the finite-start mode covariances are \citep{Bernardo:2026opv}
\begin{align}
    K_+(t,t') &= Q_+ \min(x,x'), \\
    K_-(t,t') &= \frac{Q_-\tau}{2}
    \left[
        e^{-|x-x'|/\tau} - e^{-(x+x')/\tau}
    \right], \\
    K_{+-}(t,t') &= Q_\times\tau
    \left[
        e^{-(x'-x)/\tau}\Theta(x'-x)
        + \Theta(x-x')
        - e^{-x'/\tau}
    \right],
\end{align}
where
\begin{equation}
    Q_+ = \frac{\tau_{\rm c}^2Q_{\rm c}+\tau_{\rm s}^2Q_{\rm s}}{\tau^2} \,,
    \qquad
    Q_- = Q_{\rm c}+Q_{\rm s} \,,
    \qquad
    Q_\times = \frac{\tau_{\rm c}Q_{\rm c}-\tau_{\rm s}Q_{\rm s}}{\tau} \,.
\end{equation}
The crust is the linear combination
\begin{equation}
    \Omega_{\rm c}
    = w_-\Omega_- + w_+\Omega_+ \,,
    \qquad
    w_-=\frac{\tau_{\rm s}}{\tau_{\rm c}+\tau_{\rm s}} \,,
    \qquad
    w_+=\frac{\tau}{\tau_{\rm c}+\tau_{\rm s}} \,.
\end{equation}
Hence
\begin{align}
     K_{\rm c}(t,t') = & \,
     w_-^2 K_-(t,t') \nonumber \\
     & \,\, + w_-w_+\left[K_{+-}(t,t')+K_{+-}(t',t)\right] + w_+^2 K_+(t,t') \,. \label{eq:Kc_modes}
\end{align}

For $t_i>t_j$, \eqref{eq:Kc_modes} factorizes into a rank-two semiseparable form,
\begin{equation}
\label{eq:semisep_Kc}
    K_{\rm c}(t_i,t_j) = {\bf U}_i{\bf V}_j^\top \,,
    \qquad i>j \,,
\end{equation}
where
\begin{equation}
    {\bf U}_i =
    \begin{bmatrix}
        e^{-x_i/\tau} & 1
    \end{bmatrix} \,,
\end{equation}
and
\begin{align}
    {\bf V}_j^{(1)}
    &=
    w_-^2\frac{Q_-\tau}{2}
    \left(e^{x_j/\tau}-e^{-x_j/\tau}\right)
    + w_-w_+Q_\times\tau
    \left(e^{x_j/\tau}-1\right) \,, \\
    {\bf V}_j^{(2)}
    &=
    w_+^2Q_+x_j
    + w_-w_+Q_\times\tau
    \left(1-e^{-x_j/\tau}\right) \,.
\end{align}
The diagonal is stored separately,
\begin{equation}
    A_i = K_{\rm c}(t_i,t_i)+\sigma_i^2 \,,
\end{equation}
and the upper triangle follows by symmetry. \eqref{eq:semisep_Kc} show that the dense $N\times N$ covariance does not need to be materialized.  The required information is the diagonal vector $A_i$ and the two $N\times2$ factor matrices ${\bf U}$ and ${\bf V}$.

The Gaussian log likelihood contains the two operations that normally dominate a dense GP calculation,
\begin{equation}
    \ln {\cal L}
    =
    -\frac{1}{2}
    \left[
        {\bf r}^\top {\bf C}^{-1}{\bf r}
        + \ln|{\bf C}|
        + N\ln(2\pi)
    \right],
\end{equation}
where ${\bf r}$ is the residual vector after subtracting the deterministic trend.  A dense Cholesky factorization evaluates ${\bf C}^{-1}$ and $\ln|{\bf C}|$ in ${\cal O}(N^3)$ time and stores ${\cal O}(N^2)$ numbers.

For a rank-$J$ semiseparable matrix ($J=2$), define ${\bf S}_0={\bf 0}$.  Sweeping through the data in time order gives
\begin{align}
    d_i &= A_i - {\bf U}_i{\bf S}_{i-1}{\bf U}_i^\top, \\
    {\bf W}_i &= \frac{{\bf V}_i-{\bf S}_{i-1}{\bf U}_i^\top}{d_i}, \\
    {\bf S}_i &= {\bf S}_{i-1}+d_i{\bf W}_i^\top{\bf W}_i .
\end{align}
The numbers $d_i$ are the Cholesky pivots, and therefore
\begin{equation}
    \ln|{\bf C}|=\sum_{i=1}^{N}\ln d_i .
\end{equation}
The same factorization gives the inverse action ${\bf x}={\bf C}^{-1}{\bf b}$ without forming ${\bf C}^{-1}$.  First perform the forward pass
\begin{align}
    z_i &= \frac{b_i-{\bf U}_i{\bf f}_{i-1}}{\sqrt{d_i}}, \\
    {\bf f}_i &= {\bf f}_{i-1}+{\bf W}_i^\top\sqrt{d_i}\,z_i ,
\end{align}
followed by the backward pass
\begin{align}
    x_i &= \frac{z_i}{\sqrt{d_i}}-{\bf W}_i{\bf g}_{i+1}, \\
    {\bf g}_i &= {\bf g}_{i+1}+{\bf U}_i^\top x_i .
\end{align}
The factorization costs ${\cal O}(NJ^2)$ and each solve costs ${\cal O}(NJ)$.

In the analysis we marginalize over the deterministic trend parameters with a flat prior.  Let ${\bf M}$ be the $N\times p$ trend design matrix.  The generalized least-squares estimate at fixed $\boldsymbol{\theta}$ is
\begin{equation}
    \hat{\boldsymbol{\beta}}
    =
    \left({\bf M}^\top{\bf C}^{-1}{\bf M}\right)^{-1}
    {\bf M}^\top{\bf C}^{-1}{\bf y}.
\end{equation}
The trend-marginalized log likelihood is
\begin{align}
\label{eq:trend_marginalized_semisep_likelihood}
    \ln{\cal L}(\boldsymbol{\theta})
    =
    -\frac{1}{2}
    \bigg[
        {\bf r}^\top{\bf C}^{-1}{\bf r}
        + \ln|{\bf C}|
        + \ln\left|{\bf M}^\top{\bf C}^{-1}{\bf M}\right|
        + (N-p)\ln(2\pi)
    \bigg],
\end{align}
where ${\bf r}={\bf y}-{\bf M}\hat{\boldsymbol{\beta}}$.  All appearances of ${\bf C}^{-1}$ in \eqref{eq:trend_marginalized_semisep_likelihood} are inverse actions computed by the semiseparable solve above.

This construction is inspired by \cite{2017AJ....154..220F};
which
is formulated for scalar, 1D and stationary Gaussian processes whose kernels can be written as sums of exponentials.
The neutron-star problem is intrinsically multivariate and nonstationary: the physical state is $(\Omega_{\rm c},\Omega_{\rm s})$, the eigenmodes are mixed, and the superfluid is hidden in radio data. We overcome this obstruction by using the analytical two-component solution to marginalize the latent superfluid at the covariance level \citep{Bernardo:2026opv}.  

The posterior for the stochastic parameters is obtained from
\begin{equation}
    \ln P(\boldsymbol{\theta}|{\bf y})
    =
    \ln{\cal L}(\boldsymbol{\theta})
    + \ln P(\boldsymbol{\theta})
    + {\rm Constant} \,.
\end{equation}
After the MAP or posterior samples are obtained, the hidden superfluid stochastic residual can be reconstructed by Gaussian conditioning using the analytical cross covariance between $\Omega_{\rm s}$ and $\Omega_{\rm c}$ \citep{Bernardo:2026opv}.

\section{Inference with singular model simulations}
\label{sec:singular_simulations}

\begin{table}
    \centering
    \small
    \begingroup
    \renewcommand{\arraystretch}{1.25}
    \setlength{\tabcolsep}{4pt}
    \caption{Reference parameters for the singular, nonstationary model (\ref{eq:eom_crust}--\ref{eq:xic_xis})
    and the measurement noise considered in
    Section~\ref{sec:singular_simulations}.}
    \begin{tabular}{@{}lcc@{}}
        \hline
        Parameter & Value & Unit \\
        \hline\hline
        \multicolumn{3}{@{}l}{\textit{Spin noise}} \\
        $\tau$ & $365$ & day \\
        $r$ & $0.1$ & -- \\
        $Q_{\rm c}$ & $10^{-10}$ & rad$^2$ day$^{-3}$ \\
        $Q_{\rm s}$ & $10^{-8}$ & rad$^2$ day$^{-3}$ \\
        \hline
        \multicolumn{3}{@{}l}{\textit{Measurement noise}} \\
        $R_{\rm c}$ & $10^{-9}$ & rad$^2$ day$^{-3}$ \\
        $R_{\rm s}$ & $10^{-8}$ & rad$^2$ day$^{-3}$ \\
        \hline
    \end{tabular}
    \label{tab:singular_parameters}
    \endgroup
\end{table}

Consider the singular, nonstationary model
(\ref{eq:eom_crust}--\ref{eq:xic_xis}) as a mock data-generating
process. This section provides a controlled test of the inference and reconstruction when the neutron star rotational state is fully observed, and of the systematic errors that arise when only the crust is observed.

We generate 101 evenly sampled observations over 1400 days with a cadence
$\Delta t=14\,{\rm day}$. The parameters
are shown in Table~\ref{tab:singular_parameters}.
The orders of magnitude are broadly motivated by the spin-noise parameters
inferred for PSR J1359$-$6038 \citep{ONeill:2024uiw}; however, the values
are rounded and rescaled and should not be interpreted as a new fit. We choose the composite time scale to be one year so that the
observation window contains several correlation time scales. The diffusion
coefficients are chosen to produce visibly noisy realizations over this
short observation window.

We additionally consider independent measurement noise such that the
covariance of a sampled observation is ${\bf R}\Delta t$, where
${\bf R}={\rm diag}(R_{\rm c},R_{\rm s})$. The crust measurement noise is
chosen to be comparable to the corresponding stochastic variations to produce visible error bars in the mock data.

The mock trajectories used are the stochastic residuals obtained
after removing the deterministic mean evolution. Consequently,
$\llangle\alpha_{\rm c}\rrangle$ and
$\llangle\Delta\Omega\rrangle$ specify the equivalent deterministic
parameter basis.
The posterior
analysis concerns the parameters $(\tau, r, Q_{\rm c}, Q_{\rm s})$ that determine spin wandering.

The same residual realization is analysed with the singular and minimal
models. For full-state observations, the MAP estimate based on the sufficient
statistics (Section~\ref{subsec:sufficient_statistics}) is used as an
initializer. The reported endpoint is obtained with the exact noisy
direct-GP likelihood. For crust-only observations, we use the analytical
crust covariance and reconstruct the hidden superfluid state by Gaussian
conditioning (Section~\ref{subsec:analytic_semiseparable}). The same fitted
objects are used for the reconstructions, parameter posteriors, noise
diagnostics, and evidence calculations.
Preliminary prior-expansion tests established ranges wide enough to contain the relevant likelihood support, after which those ranges were fixed across all realizations in each experiment. The purpose of this section is to illustrate the behavior of the inference and reconstruction when the data are generated from a known model.

Figure~\ref{fig:singular_reconstruction} shows the reconstructions in a
reference realization. We define the RMSE as the root mean square
difference between the posterior mean and the true state, and normalize it
by the sample standard deviation of the corresponding true state.
Pointwise inclusion fraction is the fraction of observing epochs at which the true state lies
inside the pointwise 95 percent posterior interval.

With full-state observations, the singular model follows both components
of the state. Its normalized RMSE is $0.24$ for the crust and $0.12$ for
the superfluid, with empirical 95 percent pointwise inclusion fractions
of 97 and 96 percent,
respectively. The minimal model cannot reproduce the diffusive
total angular momentum mode of the singular process. It consequently gives
a poor crust reconstruction, with a normalized RMSE of $1.10$ and only
2 percent
pointwise inclusion fraction,
even though its superfluid reconstruction follows
more of the injected path.

This behaviour is not caused by smaller superfluid error bars. The
superfluid measurement uncertainty is larger in absolute units, but the
superfluid stochastic excursions are also substantially larger and
therefore have a higher signal-to-noise ratio. In the minimal model, the
rank-one forcing ties the two components to a single anticorrelated mode.
Matching the better-resolved superfluid variations fixes the allowed crust
response, leaving the model unable to reproduce the independent diffusive
contribution present in the singular crust data.

When only crust observations are available, both models reconstruct the
1D data reasonably well. This agreement does not extend to
the hidden superfluid state. The singular reconstruction has a normalized
superfluid RMSE of $1.03$ and a median 95 percent half-width of
$6.5\times10^{-3}\,{\rm rad}\,{\rm day}^{-1}$. The minimal reconstruction
is strongly displaced from the injected path, with a normalized RMSE of
$13.0$. Its pointwise inclusion fraction is high because its uncertainty band expands to a median half-width of $1.4\times10^{-1}\,{\rm rad}\,{\rm day}^{-1}$. Pointwise inclusion fraction by such a broad band should not be confused with an informative reconstruction.

The distinction is also visible in the parameter posteriors in
Figure~\ref{fig:singular_posteriors} and Tables \ref{tab:singular_singular_posterior}-\ref{tab:singular_minimal_posterior}.

\begin{table}
    \centering
    \small
    \begingroup
    \renewcommand{\arraystretch}{1.45}
    \setlength{\tabcolsep}{1.5pt}
    \caption{Marginalized estimates obtained by fitting the singular
    model to the singular model reference realization. Units are given
    in Table~\ref{tab:singular_parameters}. `Full' and `crust' columns correspond to results with full-state and crust-only configurations, respectively. Prior entries give the lower
    and upper limits of the uniform priors.
    }
    \begin{tabular}{@{}lrrrr@{}}
        \hline
        Parameter & Input & Full & Crust & Prior \\
        \hline\hline
        $\log_{10}\tau$
        & $2.562$
        & $2.56^{+0.17}_{-0.16}$
        & $2.34^{+1.00}_{-0.78}$
        & $[1.36,3.76]$ \\

        $\log_{10}r$
        & $-1.000$
        & $-1.00^{+0.31}_{-0.30}$
        & $-1.36^{+1.06}_{-0.80}$
        & $[-2.50,0.50]$ \\

        $\log_{10}Q_{\rm c}$
        & $-10.000$
        & $-9.99^{+0.22}_{-0.22}$
        & $-9.79^{+0.20}_{-0.51}$
        & $[-13.00,-7.00]$ \\

        $\log_{10}Q_{\rm s}$
        & $-8.000$
        & $-8.00^{+0.15}_{-0.15}$
        & $-8.4^{+1.9}_{-1.6}$
        & $[-11.00,-5.00]$ \\
        \hline
    \end{tabular}
    \label{tab:singular_singular_posterior}
    \endgroup
\end{table}

\begin{table}
    \centering
    \small
    \begingroup
    \renewcommand{\arraystretch}{1.45}
    \setlength{\tabcolsep}{2pt}
    \caption{Marginalized estimates obtained by fitting the minimal model
    to the singular model reference realization. `Full' and `crust' columns correspond to results with full-state and crust-only configurations, respectively.Prior entries give the
    lower and upper limits of the uniform priors.
    }
    \begin{tabular}{@{}lrrr@{}}
        \hline
        Parameter & Full & Crust & Prior \\
        \hline\hline
        $\log_{10}\tau$
        & $4.0620^{+0.0002}_{-0.0003}$
        & $3.61^{+0.30}_{-0.33}$
        & $[1.06,4.06]$ \\

        $\log_{10}A_{\rm min}$
        & $-2.8541^{+0.0001}_{-0.0002}$
        & $-6.15^{+0.31}_{-0.29}$
        & $[-10.85,-2.85]$ \\

        $\log_{10}x_{\rm s}$
        & $-2.3982^{+0.0005}_{-0.0005}$
        & $-0.89^{+0.22}_{-0.51}$
        & $[-3.00,-0.60]$ \\
        \hline
    \end{tabular}
    \label{tab:singular_minimal_posterior}
    \endgroup
\end{table}

\begin{figure*}
    \centering
    \includegraphics[width=0.94\textwidth]
        {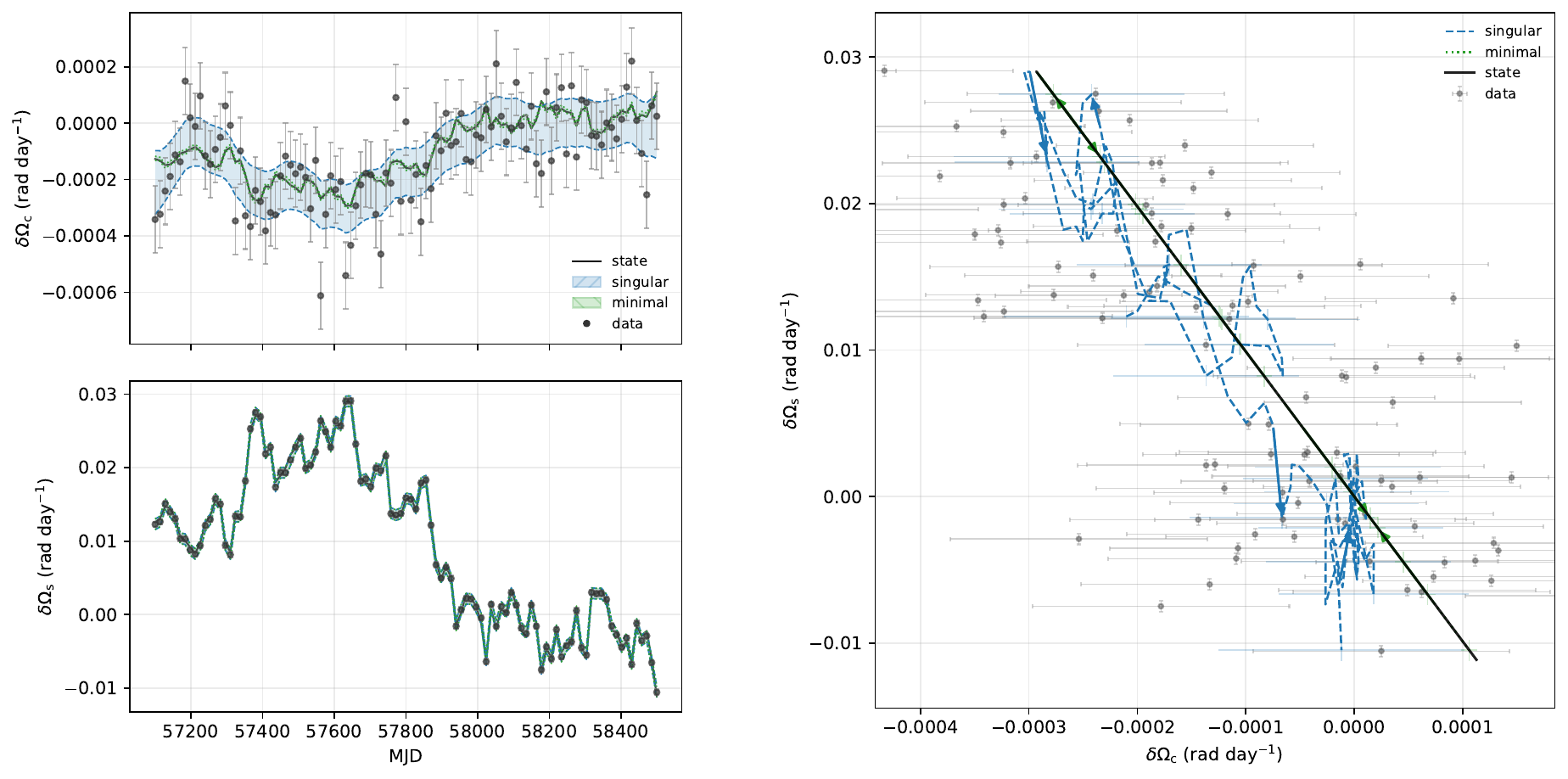}
    \includegraphics[width=0.94\textwidth]
        {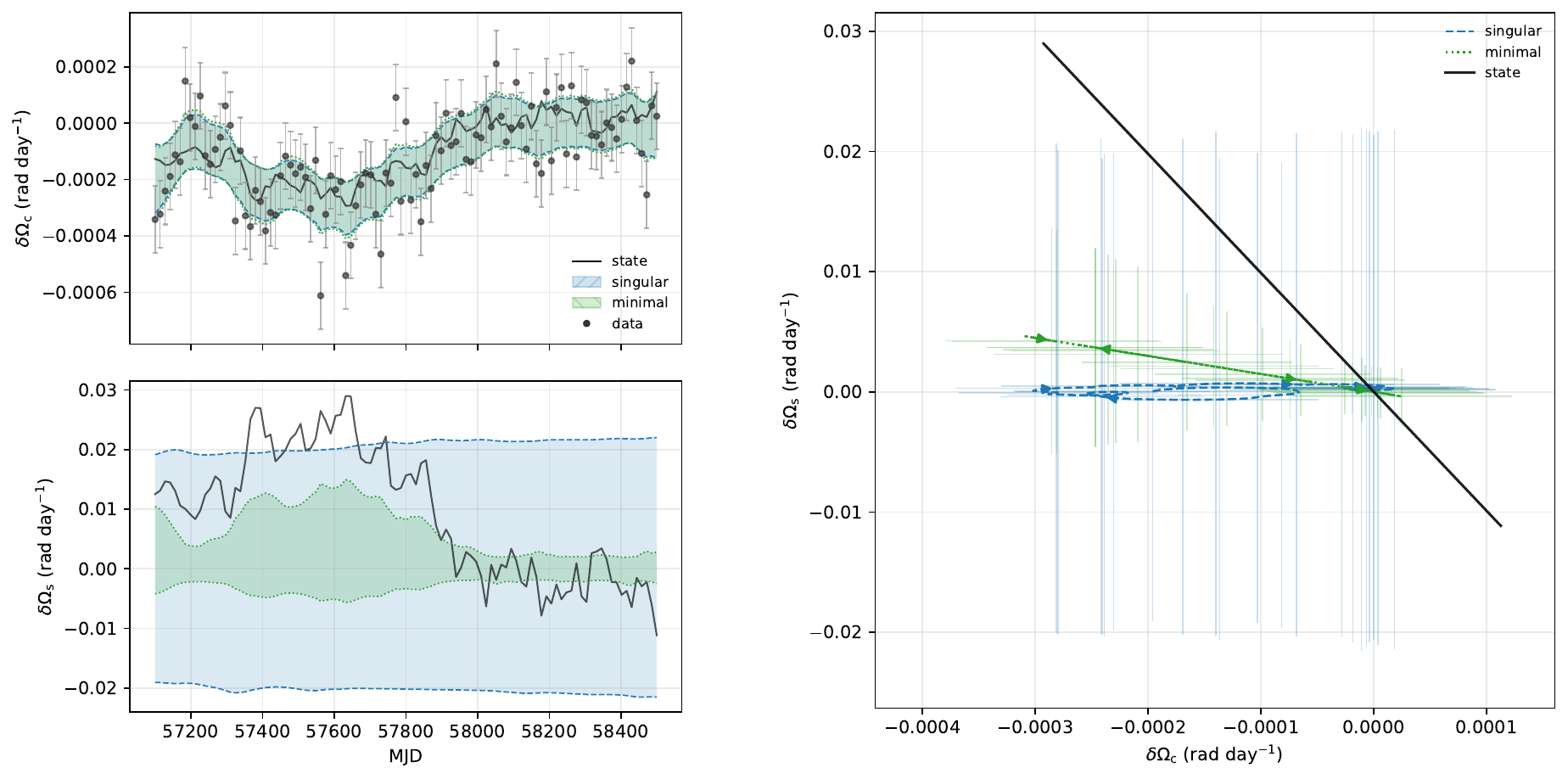}
    \caption{
    Reconstruction of a minimal-model realization using the singular
    and minimal models. The top figure uses full-state observations and
    the lower figure uses crust-only observations. The plotting
    conventions are the same as in
    Figure~\ref{fig:singular_reconstruction}.}
    \label{fig:minimal_reconstruction}
\end{figure*}

\begin{figure*}
    \centering
    \includegraphics[width=0.56\textwidth]
        {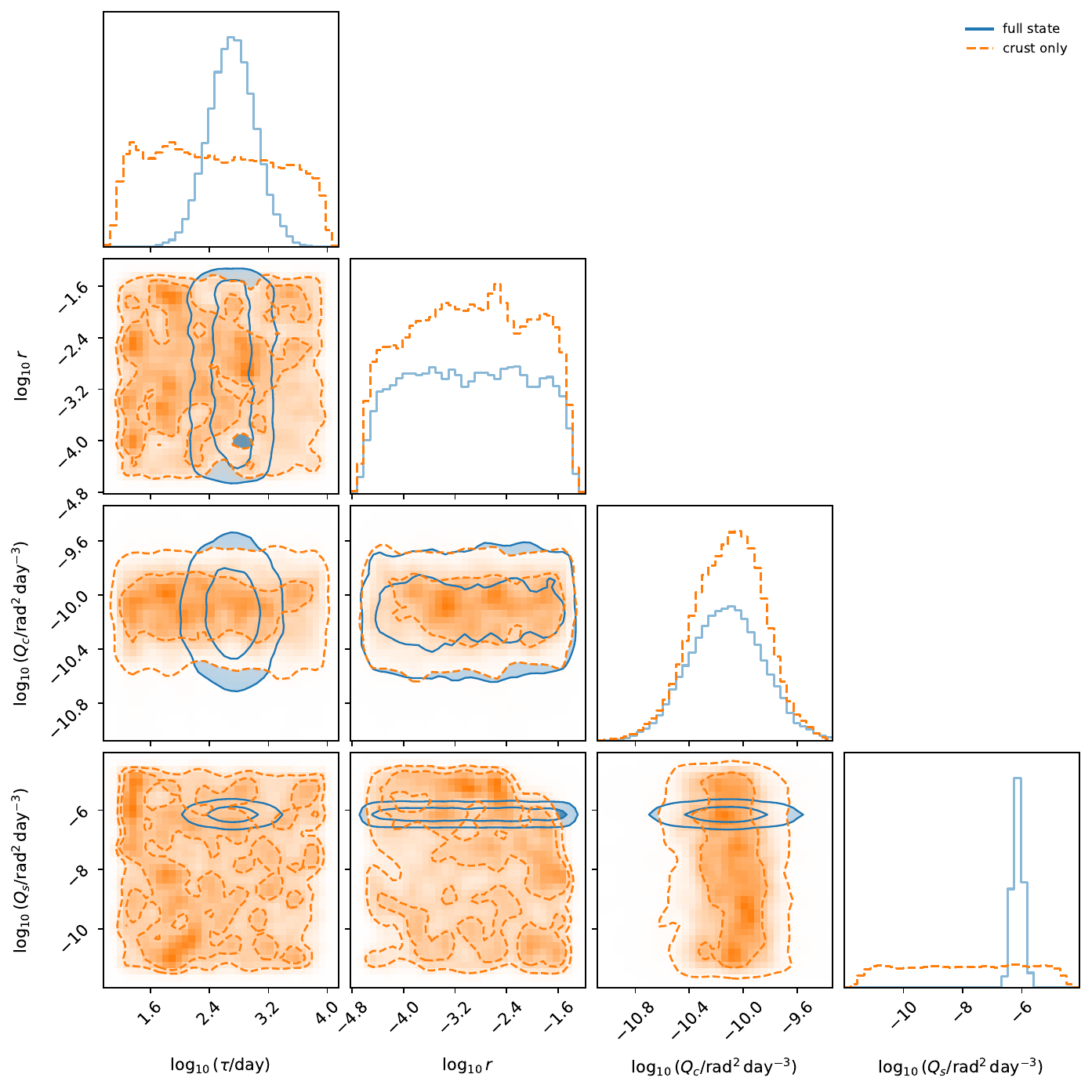}
    \hfill
    \includegraphics[width=0.41\textwidth]
        {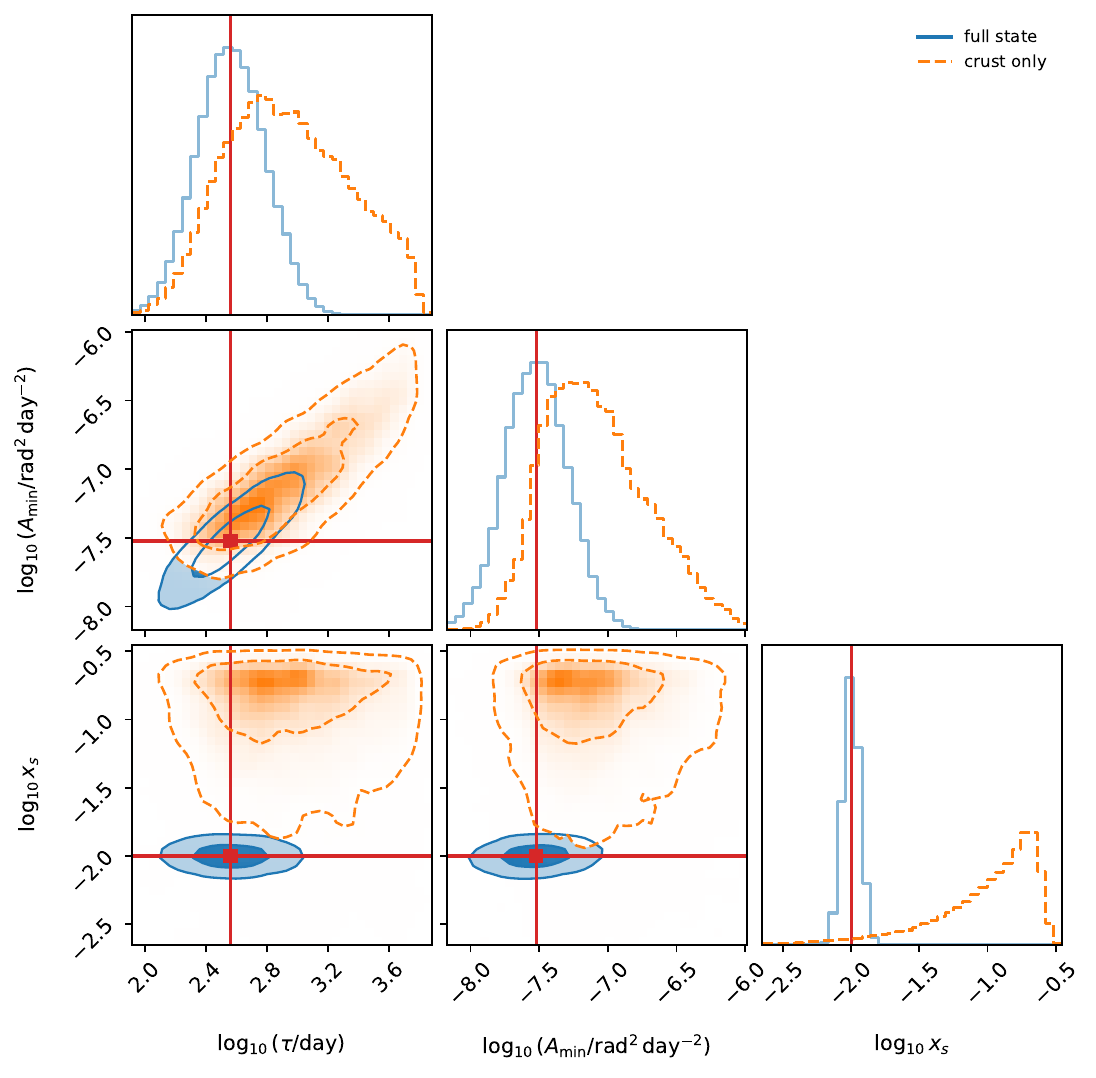}
    \caption{
    Parameter posteriors for the minimal-model realization analysed with
    the singular model (left) and the minimal model (right). Blue solid
    distributions use full-state observations and orange dashed
    distributions use crust-only observations. Red lines in the
    generating-model panel mark the injected minimal-model parameters.}
    \label{fig:minimal_posteriors}
\end{figure*}

\begin{figure*}
    \centering
    \includegraphics[width=0.88\textwidth]
        {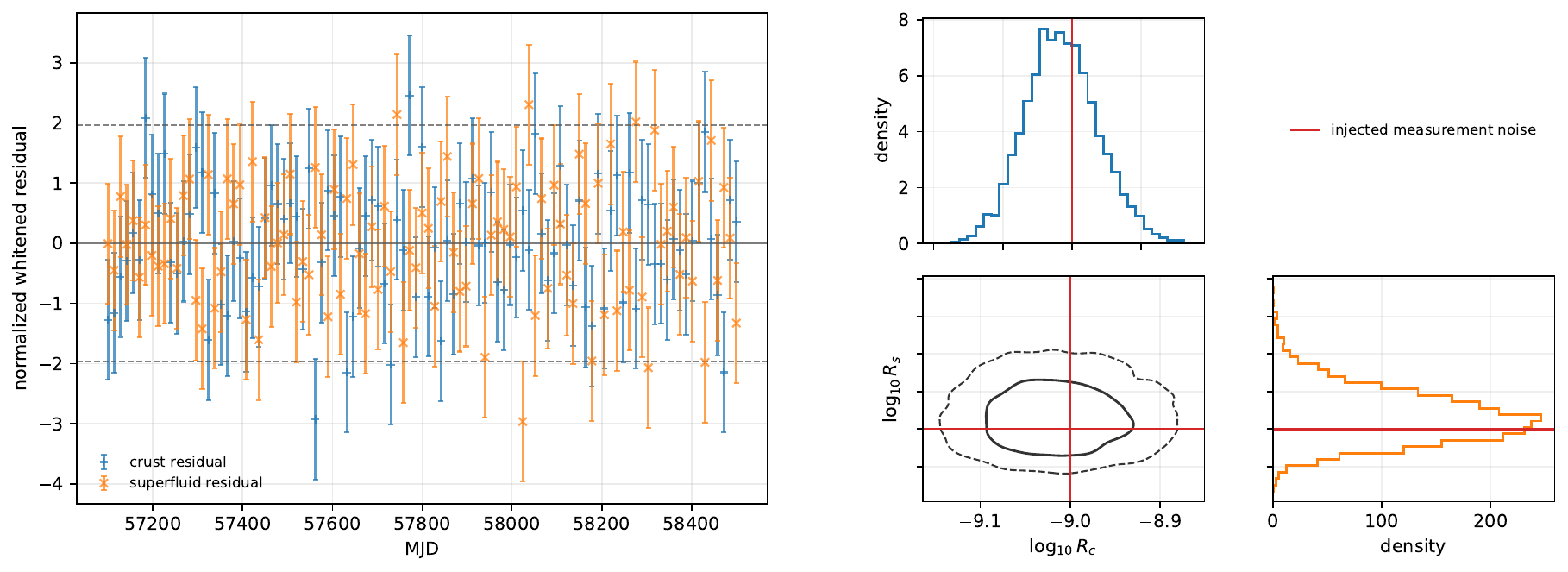}
    \includegraphics[width=0.88\textwidth]
        {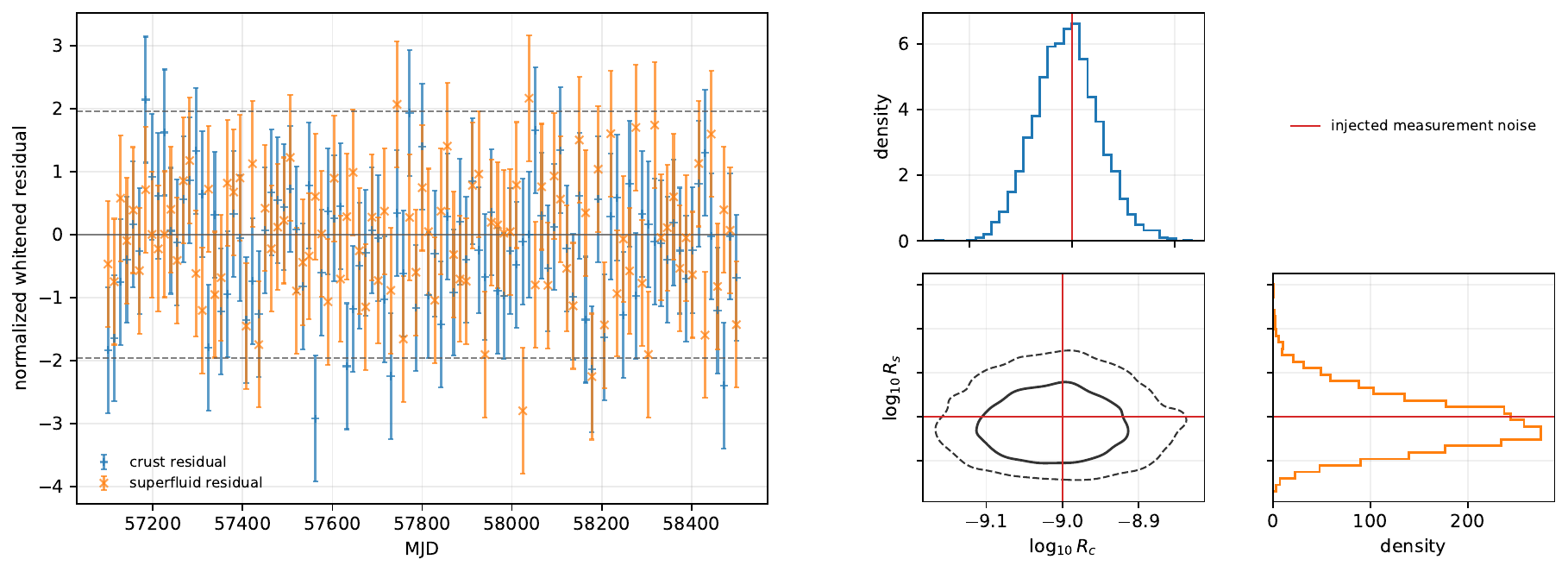}
    \includegraphics[width=0.88\textwidth]
        {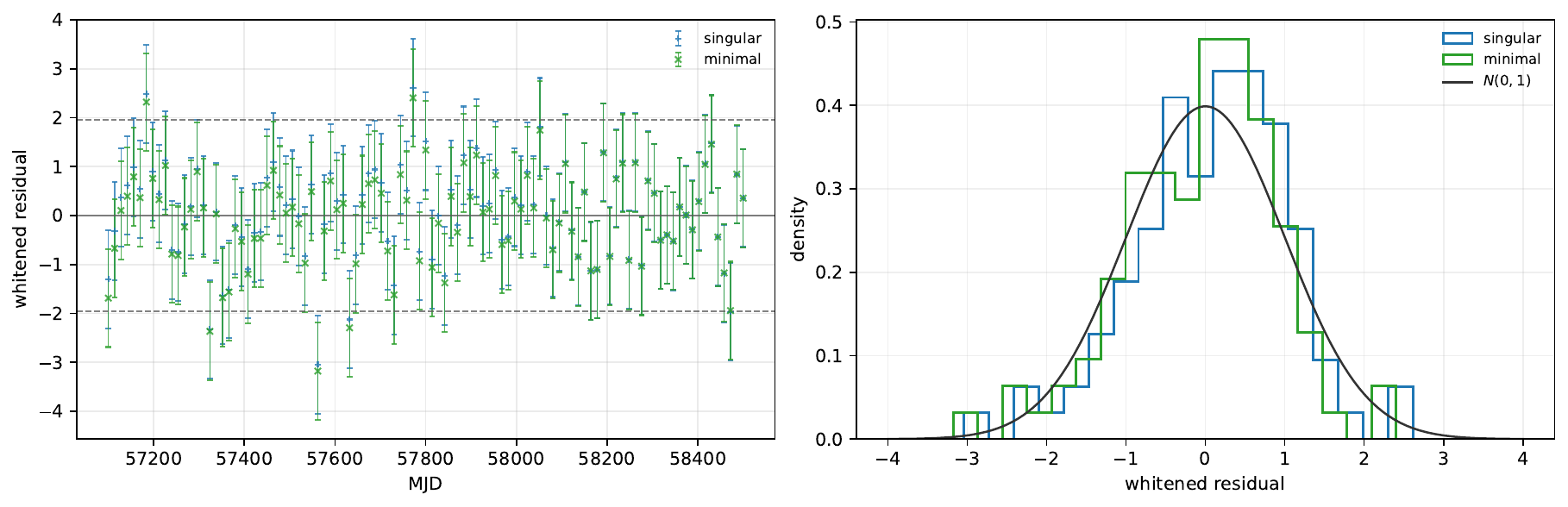}
    \caption{
    Noise diagnostics for the minimal model realization. From top to
    bottom, the figures show the full-state singular fit, the full-state
    minimal fit, and the crust-only whitened residuals. Symbols, contours,
    and injected measurement-noise lines follow
    Figure~\ref{fig:singular_noise}.}
    \label{fig:minimal_noise}
\end{figure*}

\begin{figure*}
    \centering
    \includegraphics[width=0.94\textwidth]
        {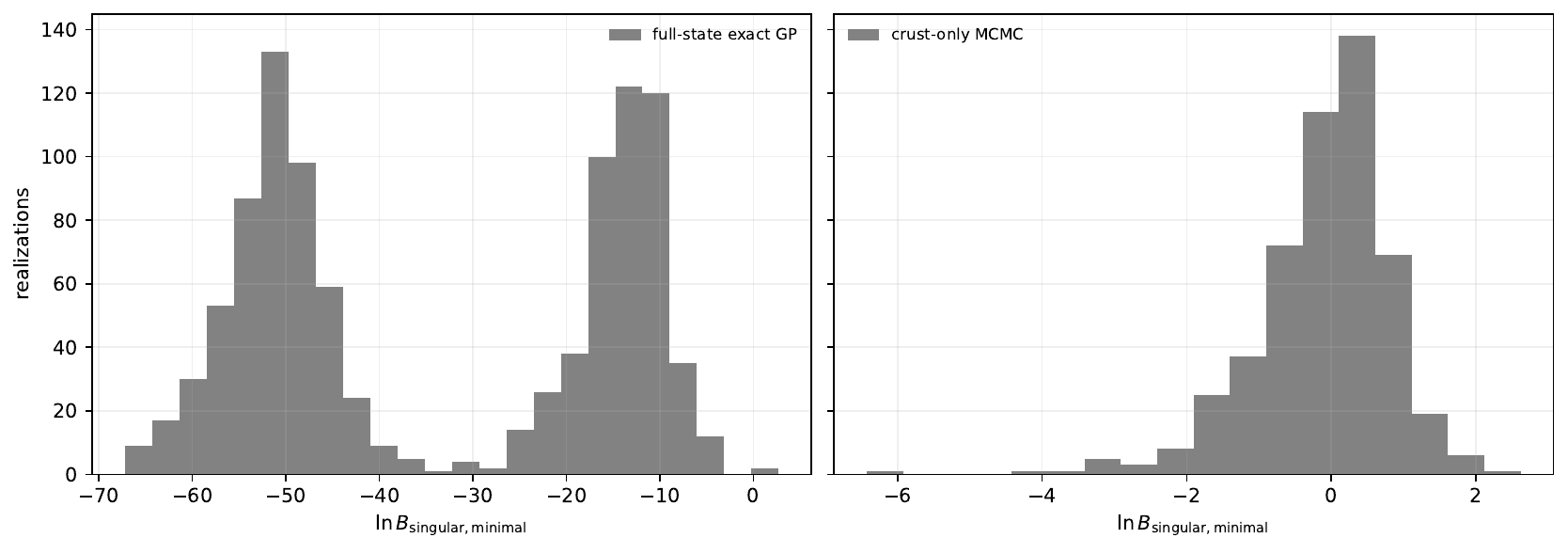}
    \caption{
    Distribution of $\ln B_{\rm singular,minimal}$ for data generated
    with the minimal model. The left panel contains 1000 full-state
    exact-GP results and the right panel contains 500 crust-only
    MCMC stepping-stone sampling results. Negative values favour the
    minimal model.}
    \label{fig:minimal_bayes_factors}
\end{figure*}

For the singular model, the full-state estimates agree with the injected parameters. The crust-only posterior is broad and strongly correlated. Although the injected values lie within much of the posterior support in this realization, the widths of the intervals show that the physical decomposition of the crust covariance is only weakly identified. For the minimal model, the localized full-state posterior is driven to the upper bounds in $\tau$ and $A_{\rm min}$. Its small marginalized uncertainties do not represent a precise measurement of parameters equivalent
to those of the singular model. They characterize a limiting configuration adopted by a misspecified model in attempting to
reproduce the singular realization.

The noise diagnostics in Figure~\ref{fig:singular_noise} support the same sentiment. For the correct full-state singular fit, the normalized crust
and superfluid residuals fluctuate without an obvious coherent trend, and
the injected $R_{\rm c}$ and $R_{\rm s}$ lie inside the corresponding
68 percent intervals. The full-state minimal fit leaves a coherent
structure in the
residuals and gives a biased estimate of
the measurement noise. The crust-only whitened residuals are
approximately consistent with a standard normal distribution under both
models.
However, a 1D residual diagnostic is not sufficient to establish that the hidden state or its physical parameters have been recovered. The displayed measurement noise distributions are residual-based consistency estimates conditional on the fitted dynamics. They include the mean posterior state variance absorbed by the smoother, but they are not jointly sampled nuisance-parameter posteriors.

We finally repeat the comparison over independent realizations. We define the log-Bayes factor
$\ln B_{\rm singular,minimal}=\ln Z_{\rm singular}-\ln Z_{\rm minimal}$, where $\ln Z_{\cal M}$ are evidences for model ${\cal M}$, such that positive values favour the singular model. Full-state evidence
is computed from the exact direct-GP likelihood using bounded Laplace
integration. Crust-only evidence is computed by MCMC stepping-stone sampling
of the analytical covariance likelihood. Local Laplace evidence is not
used for the crust-only result because its Hessian is frequently
ill-conditioned or limited by the prior boundaries. The results are shown in Figure \ref{fig:singular_bayes_factors}.

The full-state experiment correctly selects the singular model in
99.8 percent of 1000 realizations, with a 95 percent Wilson interval of
$[99.3,99.9]$ percent and median
$\ln B_{\rm singular,minimal}=806.43$. Note that the bimodality of the full-state distribution is irrelevant as both peaks (the larger one at $\sim 800$ and a smaller group at $\sim 100$) assign themselves to the positive side, preferring the generating model. The crust-only experiment selects
the singular model in 48.8 percent of 500 realizations, with interval
$[44.4,53.2]$ percent and median
$\ln B_{\rm singular,minimal}=-0.05$. Changing the width of the singular
prior by one decade also changes the sign or magnitude of the crust-only
Bayes factor. The crust-only evidence is both data-limited and
prior-limited and does not reliably distinguish the two models. This also supports why the crust-only evidence does not penalize the model with the larger parameter space.\footnote{When comparing two models, the Occam argument relies on a few assumptions: that both have the same likelihood and all their parameters constrained, such that the model with the larger parameter space will occupy a smaller fraction of its prior volume.}

The singular model experiment separates two questions that can appear
similar when only one time series is considered. The observed crust can
be reconstructed accurately from crust-only data, but the decomposition of its covariance into the underlying two-component dynamics is not robust. The hidden superfluid prediction, physical parameter estimates, and model evidence all become sensitive to model assumptions and prior volume once the second state is unavailable observationally.

\section{Inference with minimal model simulations}
\label{sec:minimal_simulations}

\begin{table}
    \centering
    \small
    \begingroup
    \renewcommand{\arraystretch}{1.25}
    \setlength{\tabcolsep}{4pt}
    \caption{Reference parameters for the minimal model (\ref{eq:Jeom_minimal}-\ref{eq:wncov_minimal}) and the
    measurement noise considered in Section~\ref{sec:minimal_simulations}.}
    \begin{tabular}{@{}lcc@{}}
        \hline
        Parameter & Value & Unit \\
        \hline\hline
        \multicolumn{3}{@{}l}{\textit{Spin noise}} \\
        $\tau$ & $365$ & day \\
        $A_{\rm min}$ & $3\times10^{-8}$ &
            rad$^2$ day$^{-2}$ \\
        $x_{\rm s}$ & $10^{-2}$ & -- \\
        \hline
        \multicolumn{3}{@{}l}{\textit{Measurement noise}} \\
        $R_{\rm c}$ & $10^{-9}$ & rad$^2$ day$^{-3}$ \\
        $R_{\rm s}$ & $10^{-8}$ & rad$^2$ day$^{-3}$ \\
        \hline
    \end{tabular}
    \label{tab:minimal_parameters}
    \endgroup
\end{table}

This section continues the experiment of Section \ref{sec:singular_simulations} with the minimal, stationary model (\ref{eq:Jeom_minimal}-\ref{eq:wncov_minimal}) as the mock
data-generating process. The observation window, cadence, number of samples, and measurement noise are kept unchanged. The methods used and presentation are also the same.

The parameters are shown in Table~\ref{tab:minimal_parameters}. The one-year relaxation time
allows the stationary mode to evolve appreciably over the 1400-day
observation. The choices of $A_{\rm min}$ and $x_{\rm s}$ give
characteristic crust and superfluid amplitudes of
$1.73\times10^{-4}\,{\rm rad}\,{\rm day}^{-1}$ and
$1.71\times10^{-2}\,{\rm rad}\,{\rm day}^{-1}$, respectively. The crust
signal is comparable to its per-sample measurement uncertainty, whereas the superfluid state would be measured at high signal-to-noise if it were directly observed. As in the singular model experiment, we simulate and analyze the stochastic residual after removal of the deterministic mean.
The spin noise parameters are $\tau, x_{\rm s}, A_{\rm min}$. Following Section \ref{sec:singular_simulations}, preliminary prior-expansion tests were made to identify ranges wide enough that contain the relevant likelihood support to illustrate inference and reconstruction with mock data.

Figure~\ref{fig:minimal_reconstruction} shows the reconstructions of a
reference realization. The definitions of normalized RMSE and pointwise inclusion fraction
are the same as in Section~\ref{sec:singular_simulations}. With full-state
observations, both models follow the measured paths. The correct minimal
fit has normalized RMSEs of $0.034$ for both components and 97 percent
pointwise inclusion fraction. The singular model obtains a similarly accurate reconstruction
despite its additional diffusive mode. This does not make the models
equivalent. Independent singular model forcing cannot exactly reproduce the rank-one anticorrelated forcing of the minimal process, but its additional parameters allow it to approximate one finite full-state
realization.

With crust-only observations, both models reconstruct the measured crust
with approximately 98 percent pointwise inclusion fraction. However, neither model gives a
robust prediction of the hidden superfluid path. The singular model has a
normalized superfluid RMSE of $1.32$ and a median 95 percent half-width
of $2.1\times10^{-2}\,{\rm rad}\,{\rm day}^{-1}$. The minimal model has a
normalized RMSE of $1.11$ and a narrower median half-width of
$3.8\times10^{-3}\,{\rm rad}\,{\rm day}^{-1}$, but its empirical
95 percent pointwise inclusion fraction is only 16.8 percent. In this realization, the
correct model is overconfident about the hidden state. The failure is not caused by poor interpolation of the observed crust, but by
weak identification of the superfluid reservoir fraction $x_{\rm s}$ that converts crust fluctuations into a superfluid prediction.

The marginalized estimates are presented in Figure \ref{fig:minimal_posteriors} and Tables \ref{tab:minimal_minimal_posterior}-\ref{tab:minimal_singular_posterior}.

\begin{table}
    \centering
    \small
    \begingroup
    \renewcommand{\arraystretch}{1.45}
    \setlength{\tabcolsep}{1.5pt}
    \caption{Marginalized estimates obtained by fitting the minimal model
    to the minimal model reference realization. `Full' and `crust' columns correspond to results with full-state and crust-only configurations, respectively. Prior entries give the
    lower and upper limits of the uniform priors.
    }
    \begin{tabular}{@{}lrrrr@{}}
        \hline
        Parameter & Input & Full & Crust & Prior \\
        \hline\hline
        $\log_{10}\tau$
        & $2.562$
        & $2.56^{+0.22}_{-0.22}$
        & $2.91^{+0.46}_{-0.39}$
        & $[1.06,4.06]$ \\

        $\log_{10}A_{\rm min}$
        & $-7.523$
        & $-7.52^{+0.23}_{-0.22}$
        & $-7.13^{+0.44}_{-0.32}$
        & $[-10.85,-2.85]$ \\

        $\log_{10}x_{\rm s}$
        & $-2.000$
        & $-2.000^{+0.038}_{-0.037}$
        & $-0.91^{+0.23}_{-0.48}$
        & $[-3.00,-0.60]$ \\
        \hline
    \end{tabular}
    \label{tab:minimal_minimal_posterior}
    \endgroup
\end{table}

\begin{table}
    \centering
    \small
    \begingroup
    \renewcommand{\arraystretch}{1.45}
    \setlength{\tabcolsep}{2pt}
    \caption{Marginalized estimates obtained by fitting the singular
    model to the minimal model reference realization. `Full' and `crust' columns correspond to results with full-state and crust-only configurations, respectively. Prior entries give
    the lower and upper limits of the uniform priors.}
    \begin{tabular}{@{}lrrr@{}}
        \hline
        Parameter & Full & Crust & Prior \\
        \hline\hline
        $\log_{10}\tau$
        & $2.71^{+0.30}_{-0.30}$
        & $2.46^{+0.99}_{-0.90}$
        & $[1.16,3.96]$ \\

        $\log_{10}r$
        & $-3.0^{+1.1}_{-1.1}$
        & $-3.0^{+1.0}_{-1.0}$
        & $[-4.60,-1.40]$ \\

        $\log_{10}Q_{\rm c}$
        & $-10.12^{+0.26}_{-0.27}$
        & $-10.10^{+0.24}_{-0.28}$
        & $[-13.50,-6.50]$ \\

        $\log_{10}Q_{\rm s}$
        & $-6.133^{+0.061}_{-0.063}$
        & $-8.0^{+2.3}_{-2.4}$
        & $[-11.50,-4.50]$ \\
        \hline
    \end{tabular}
    \label{tab:minimal_singular_posterior}
    \endgroup
\end{table}

The full-state estimates agree with all three spin-noise parameters. The crust-only covariance continues to constrain the amplitude
and relaxation time scale, but it does not independently determine how the
fluctuations are partitioned between the two components. Consequently, the inferred
$x_{\rm s}$ is displaced from the injected value, leading directly to the biased superfluid reconstruction.
On the other hand,
the singular fit can closely reproduce the observed full-state path, but
its inferred parameters are not measurements of minimal model
quantities. In the crust-only case, the broad distributions again show
that several singular model parameter combinations approximate the 1D crust covariance.

Figure~\ref{fig:minimal_noise} shows that the full-state residuals and
measurement noise estimates are well behaved for both models. The
injected $R_{\rm c}$ and $R_{\rm s}$ lie inside the residual-based 68 percent intervals of the singular and minimal fits. The crust-only
whitened residuals are likewise approximately standard normal under both
models. Residual whiteness tests whether a model describes the observed
covariance, but does not guarantee that the latent physical decomposition
or the hidden superfluid state has been recovered.

Figure \ref{fig:minimal_bayes_factors} shows the evidences for full-state and crust-only experiments with the minimal model-generated data. The evidence ensemble distinguishes the models when both states are
observed. As with the previous section, we do not pay attention to the bimodality of the Bayes factor as both peaks (around $-50$ and $-13$) provide support for the data-generating model. The full-state analysis correctly selects the minimal model in
99.8 percent of 1000 realizations, with a 95 percent Wilson interval of
$[99.3,99.9]$ percent and median
$\ln B_{\rm singular,minimal}=-42.71$. The evidence therefore penalizes
the additional singular model freedom even when that model closely
reconstructs one finite full-state path.

The crust-only MCMC stepping-stone evidence selects the minimal
model in only 47.8 percent of 500 realizations, with interval
$[43.5,52.2]$ percent. Its median Bayes factor is
$\ln B_{\rm singular,minimal}=0.05$, consistent with no preference. As in the singular model experiment, changing the singular prior width by one decade materially changes the sign or magnitude of the Bayes factor.

The two simulation experiments give the same central result from opposite
directions. Full-state observations recover the parameters of the generating process, reconstruct both angular velocities, and distinguish the stationary and nonstationary models reliably. Crust-only observations can still give an accurate reconstruction of the measured crust and satisfactory
whitened residuals. They do not reliably provide the same information about the hidden superfluid, the physical partitioning of the stochastic forcing,
or the model evidence. The dominant limitation is not the
interpolation of the radio-observed state, but the systematic uncertainty introduced by projecting a 2D stochastic process onto a single observed component.

The singular (\ref{eq:eom_crust}--\ref{eq:xic_xis}) and minimal (\ref{eq:Jeom_minimal}--\ref{eq:wncov_minimal}) models have different parameterizations and are assigned separate normalized priors, whose integration supplies the corresponding Occam penalty. The prior ranges were established through preliminary expansion tests and held fixed within each simulation ensemble. The singular-model prior differs between the two experiments (Sections \ref{sec:singular_simulations}-\ref{sec:minimal_simulations}) because its high-likelihood region shifts when it is fitted as a misspecified approximation to minimal-model data. The reported selection fractions are consequently conditional on their respective model-prior pairs rather than a calibration of a single universal classifier. This distinction does not affect the strong full-state preference. The crust-only Bayes factors remain sensitive to the tested prior ranges.

\section{Conclusions}
\label{sec:conclusions}

Two-component models of pulsar spin wandering, like their use in explaining glitches, are a natural observational starting point for associating radio timing noise with underlying physical processes in neutron stars. However, these models are completely determined by the rotational states of both the crust and the superfluid. For the observing span, cadence, measurement noise, and model-prior pairs considered here, crust-only evidence does not reliably identify the generating model and remains sensitive to prior volume. Observing only a single component does not provide sufficient information to reliably constrain the model or to make robust predictions, whether of future configurations or of a radio-hidden state \citep{Meyers:2021myb, Antonelli:2022gqw}. This work has provided an explicit, quantitative example of this limitation.

We worked with two spin noise or spin wandering models, distinguished by the dynamics of their total angular momentum: a singular, nonstationary model (\ref{eq:eom_crust}--\ref{eq:xic_xis}) and a minimal, stationary one (\ref{eq:Jeom_minimal}-\ref{eq:wncov_minimal}). The former produces nonstationary spin noise owing to a diffusive total angular momentum mode \citep{Meyers:2021myb}, while the latter produces stationary spin noise anchored on a conserved total angular momentum \citep{Antonelli:2022gqw}. Both models were written as a shifted 2D OU process, drawing on their analogy with 2D Brownian motion to develop fast and scalable Gaussian process methods for inference and prediction (Section \ref{sec:fast_bayesian}). With simulated mock data based on both models, we have shown that the two are fully constrainable and reliably distinguishable when both components are observed, but that robust inferences and predictions cannot be obtained when only a single component is observationally accessible (Sections \ref{sec:singular_simulations}-\ref{sec:minimal_simulations}). This is analogous to constraining a 2D random walk that is observed only along a single axis.

Beyond these physical conclusions, this work is fundamentally a methods contribution. The Gaussian process models constructed here are physical, derived directly from the two-component equations of motion rather than from a generic phenomenological kernel or power spectral density, which is what makes them testable and falsifiable and lets the stationarity of spin noise be read directly as a statement about the conservation of a neutron star's total angular momentum. They are also scalable, with cost growing linearly rather than cubically with the number of observations, a necessity for the long and ever-growing datasets of pulsar timing. In this sense, the paper pushes the modelling of timing noise in a more physical direction.

In radio pulsar timing, the observed pulsar phase and its associated timing noise are tied to the crust's rotational state. The explanation of glitches by a two-component model provides strong evidence for a neutron superfluid, but glitches constrain only part of the two-component parameter space and are typically treated as a separate phenomenon from timing noise. The results of this work indicate that spin wandering in the two-component model may only be robustly constrained with simultaneous, independent observations of the rotational states of the crust and the superfluid, or with an observation of the full state space of the neutron star. Continuous gravitational wave observations tied to the neutron star core, and possibly to the interior superfluid motion, are a promising avenue to complement radio timing and access this state space. The methods developed here are directly applicable to such joint analyses, and we intend to apply them to real pulsar timing data, expanding the methods developed here at the level of angular velocities, to phases, in a different work.

\section*{Acknowledgements}
The author is grateful to Reinhard Prix for valuable discussion on neutron star hydrodynamics and continuous gravitational waves, Boris Goncharov and Wang-Wei Yu for comments on a preliminary draft, and Bruce Allen for discussions that shaped the direction of this work.


\appendix

\section{Schematic coarse-graining of a two-fluid model}
\label{sec:two_fluid}

We provide a sketch of the derivation of the body-averaged two-component model starting with a two-fluid description of a neutron star \citep{Prix:1999gk,Prix:2002jn}.
For clarity, we neglect entrainment, composition-changing reactions, elasticity, and magnetic stresses, and treat the stellar structure and moments of inertia as fixed.

Let $x\in\{{\rm c},{\rm s}\}$ label the charged/crustal component and the neutron superfluid component. We start from the Newtonian two-fluid equations,
\begin{align}
    \partial_t \rho_x + \nabla \cdot (\rho_x {\bf v}_x) &= 0 \,,
    \label{eq:app_continuity}
    \\
    \rho_x \left( \partial_t + {\bf v}_x \cdot \nabla \right) {\bf v}_x
    &= - \nabla P_x - \rho_x \nabla \Phi
    + {\bf f}^{\rm int}_x
    + {\bf f}^{\rm ext}_x
    + \delta {\bf f}_x \,
    \label{eq:app_euler} \\
    \nabla^2 \Phi &= 4\pi G_{\rm N} ( \rho_{\rm c} + \rho_{\rm s} ) \,.
    \label{eq:poisson}
\end{align}
${\bf f}^{\rm int}_x$, ${\bf f}^{\rm ext}_x$, and $\delta{\bf f}_x$ are the internal, external, and unresolved fluctuating force densities acting on component $x$, respectively.

Define the $z$-component of angular momentum of each component by
\begin{equation}
    J_x = \int d^3r \, \rho_x \left( {\bf r} \times {\bf v}_x \right)_z \, .
    \label{eq:app_Jx_def}
\end{equation}
Taking the angular-momentum moment of \eqref{eq:app_euler}, and using \eqref{eq:app_continuity},\footnote{The continuity equation gives
$\rho_x D_t {\bf v}_x = \partial_t (\rho_x {\bf v}_x) + \nabla \cdot (\rho_x {\bf v}_x {\bf v}_x)$
where $D_t = \partial_t + {\bf v}_x \cdot \nabla$ is the convective derivative;
$\rho_x {\bf v}_x$ and $\rho_x {\bf v}_x ^2 / 2$
are the fluid $x$'s momentum and kinetic energy densities, respectively.} gives
\begin{equation}
    \dot{J}_x
    =
    T^{\rm int}_x
    + N_x
    + \xi_x(t) \, ,
    \label{eq:app_Jx_balance}
\end{equation}
where
\begin{align}
    T^{\rm int}_x
    &= \int d^3r \, \left( {\bf r} \times {\bf f}^{\rm int}_x \right)_z \, ,
    \\
    N_x
    &= \int d^3r \, \left( {\bf r} \times {\bf f}^{\rm ext}_x \right)_z \, ,
    \\
    \xi_x(t)
    &= \int d^3r \, \left( {\bf r} \times \delta{\bf f}_x \right)_z \, .
\end{align}
In deriving \eqref{eq:app_Jx_balance} we assume that pressure and gravity exert no net torque about the rotation axis. This is appropriate for an isolated, nearly axisymmetric star in the slow-rotation approximation.

We impose a rigid-rotation ansatz for each component,
\begin{equation}
    {\bf v}_x(t,{\bf r}) = \Omega_x(t) \, \hat{\bf z} \times {\bf r} \, ,
    \label{eq:app_rigid_ansatz}
\end{equation}
so that
\begin{equation}
    J_x = I_x \Omega_x \, ,
    \qquad
    I_x = \int d^3r \, \rho_x r_\perp^2 \, .
\end{equation}
The body-averaged equations become
\begin{align}
    I_{\rm c}\dot{\Omega}_{\rm c}
    &= T^{\rm int}_{\rm c}
    + N_{\rm c}
    + \xi_{\rm c}(t) \, ,
    \label{eq:app_body_c}
    \\
    I_{\rm s}\dot{\Omega}_{\rm s}
    &= T^{\rm int}_{\rm s}
    + N_{\rm s}
    + \xi_{\rm s}(t) \, .
    \label{eq:app_body_s}
\end{align}
For small differential rotation, the internal torque may be expanded to leading order in the lag \citep{Andersson:2005rs, Haskell:2017lkl}:
\begin{equation}
    \Omega_- \equiv \Omega_{\rm c} - \Omega_{\rm s} \, .
\end{equation}
The linear-response form is
\begin{align}
    T^{\rm int}_{\rm c} &= - K \Omega_- \, ,
    \\
    T^{\rm int}_{\rm s} &= K \Omega_- \, ,
\end{align}
where the coefficient $K$ encodes the strength of the effective mutual friction. The opposite signs express conservation of total
angular momentum by the internal coupling. Writing down
\begin{equation}
    K = \frac{I_{\rm c}}{\tau_{\rm c}} = \frac{I_{\rm s}}{\tau_{\rm s}} \, ,
\end{equation}
we obtain (\ref{eq:eom_crust}--\ref{eq:eom_superfluid}).

Analogous to Brownian motion, the dynamics of a neutron star relevant to pulsar timing is dominated by a few body-averaged variables; the rest of the hydrodynamic degrees of freedom are unresolved. The resolved variables are the body-averaged angular velocities $(\Omega_{\rm c},\Omega_{\rm s})$. The unresolved variables include vortex motion, vortex pinning and unpinning, turbulent fluctuations, crustal elasticity, magnetospheric variability, and other short-scale hydrodynamic degrees of freedom. Projecting out these unresolved variables gives, in general, a non-Markov stochastic equation of the schematic form
\begin{equation}
    I_x \dot{\Omega}_x(t)
    =
    N_x
    + T^{\rm int}_x[\Omega_{\rm c},\Omega_{\rm s}]
    - \sum_y \int_{-\infty}^{t} dt' \,
    K_{xy}(t-t') \Omega_y(t')
    + \xi_x(t) \, .
    \label{eq:app_generalized_langevin}
\end{equation}
Projection-operator methods show that eliminating unresolved degrees of freedom generally yields an effective equation containing an instantaneous drift, a memory term, and an orthogonal fluctuating force \citep{1965PThPh..33..423M,2000PNAS...97.2968C}.
The two-component spin-wandering model corresponds to the Markov, linearized limit of \eqref{eq:app_generalized_langevin}, in which the memory kernels are approximated by local damping terms and the fluctuating torques by Gaussian white noise;
\begin{equation}
    \langle \xi_x(t) \rangle = 0 \, ,
    \qquad
    \langle \xi_x(t)\xi_y(t') \rangle
    =
    D_{xy}\delta(t-t') \, .
    \label{eq:app_noise_general}
\end{equation}
For purely internal fluctuations, the total angular momentum should not receive a stochastic impulse, and one should impose
\begin{equation}
    \xi_{\rm c}(t) + \xi_{\rm s}(t) = 0 \,,
    \label{eq:app_internal_noise}
\end{equation}
or write down
\begin{equation}
    \xi_{\rm c}(t)=\xi_{\rm ext}(t)-\xi_{\rm int}(t) \, ,
    \qquad
    \xi_{\rm s}(t)=\xi_{\rm int}(t) \, ,
    \label{eq:app_noise_decomp}
\end{equation}
where $\xi_{\rm ext}(t)$ represents external torque noise, e.g., magnetospheric or accretion variability \citep{Meyers:2021myb,Meyers:2021slh}, and $\xi_{\rm int}(t)$ represents stochastic internal angular-momentum exchange. This decomposition makes explicit which part of the noise drives the total angular momentum and which part only excites the crust-superfluid lag \citep{Antonelli:2022gqw, Haskell:2017lkl}.

\section{Fluctuation-dissipation theorem}
\label{sec:fluctuation_dissipation_theorem}

In Brownian motion, above formally solving the stochastic problem by deriving the probability distribution of the random variables, it is the realization that the damping term and the noise term are related that is physically meaningful. This fluctuation-dissipation relation is the reflection of microscopic dynamics in thermal equilibrium and was key to understanding that the random motion exhibited by a pollen grain in a fluid is due to the molecular collisions with the fluid molecules. Through this, the observation of Brownian motion provided a compelling evidence for the molecular composition of matter \citep{1905AnP...322..549E, Langevin1908, 10.1119/1.18725, Uhlenbeck:1930zz, Chandrasekhar:1943ws}.

The standard 3D Brownian motion can be described by the Langevin equation with the drift matrix $\mathbf{D} = \lambda \mathbf{I}$ and the diffusion matrix $\mathbf{Q} = \Gamma \mathbf{I}$, where $\lambda$ is the friction coefficient and $\Gamma$ is the strength of the noise correlation function. The fluctuation-dissipation relation is $\Gamma = 2 k_B T \lambda / m$, where $k_B$ is the Boltzmann constant, $T$ is the temperature of the fluid, and $m$ is the mass of the particle. This relation ensures that the system reaches thermal equilibrium with a Maxwell-Boltzmann distribution at temperature $T$.

Thermal noise in an RC circuit has also been explained by the fluctuation-dissipation theorem \citep{1928PhRv...32...97J, 1928PhRv...32..110N}. Applying Kirchhoff's law, the voltage across the resistor $R$ and the capacitor $C$ is given by $V = -R I + Q/C$ where $Q$ is the charge and $I=dQ/dt$ is the current through the circuit. The thermal noise voltage can be modeled as a stochastic process with zero mean and a correlation function given by $\langle V(t) V(t') \rangle = 2 k_B T R \delta(t - t')$. The fluctuation-dissipation theorem relates the existence of thermal noise in a circuit to the dissipative properties of the resistor and is a manifestation of the microscopic dynamics of electrons in the resistor. 

The same microscopic dynamics that gives rise to dissipation also gives rise to fluctuations \citep{1951PhRv...83...34C}. The fluctuation-dissipation theorem is a fundamental principle with wide ranging applications that connects the response of a system to external perturbations with the internal fluctuations of the system in thermal equilibrium. This may also be viewed as a balance check for macroscopic dynamics: without dissipation the fluctuations would grow without bound and the system would not be stable. In thermal equilibrium, fluctuation and dissipation can be viewed as two sides of the same coin and the observation of one informs the other.

\section{Two-dimensional Brownian motion}
\label{sec:2d_brownian}

\begin{figure}
\centering
\includegraphics[width=0.45\textwidth]{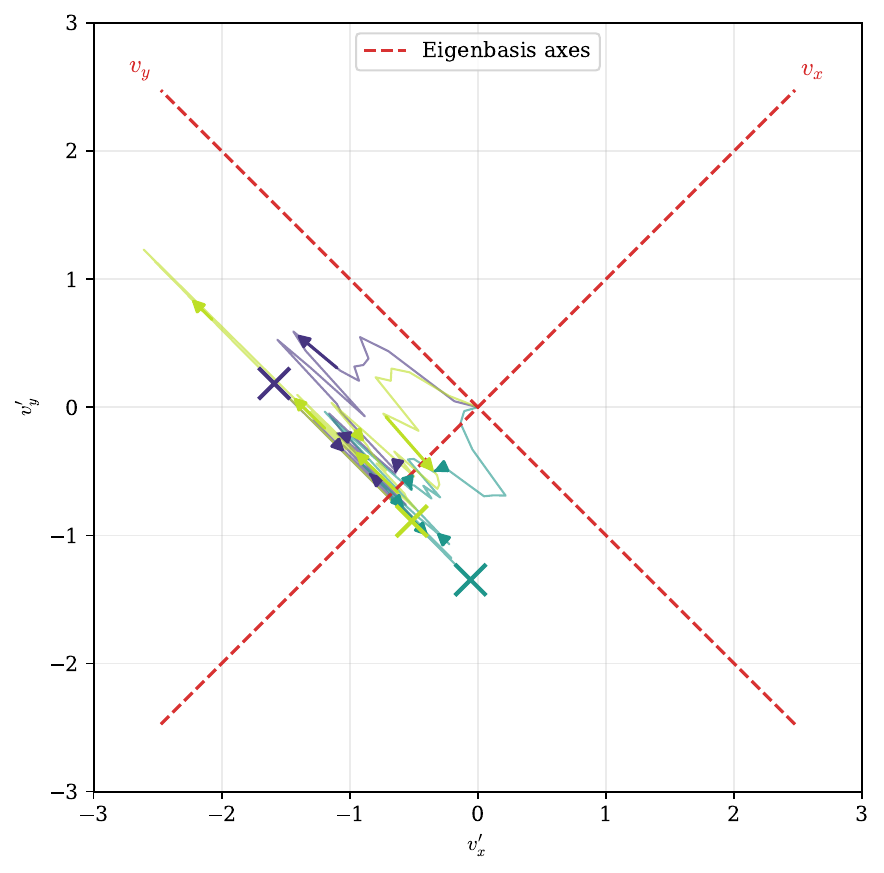}
\caption{Three realizations of 2D singular Brownian motion with $(\lambda_x, \sigma_y, f_x)=(1,1,-1)$ for $t \in [0, 5 \tau_x]$, zero initial velocity, and $\theta=45^\circ$. Time is expressed in units of the relaxation time of the OU eigenmode $\tau_x=\lambda_x^{-1}$. Crosses mark the position of the walker at $t = 5 \tau_x$. Random motion is damped in $v_x$ and diffusive along $v_y$. Observed processes $(v_x', v_y')$ are a superposition of $(v_x, v_y)$ eigenmodes. The state approaches a steady state value $\llangle v_x \rrangle = f_x / \lambda_x$.}
\label{fig:brownian_realizations}
\end{figure}

\begin{figure*}
\centering
\includegraphics[width=0.8\textwidth]{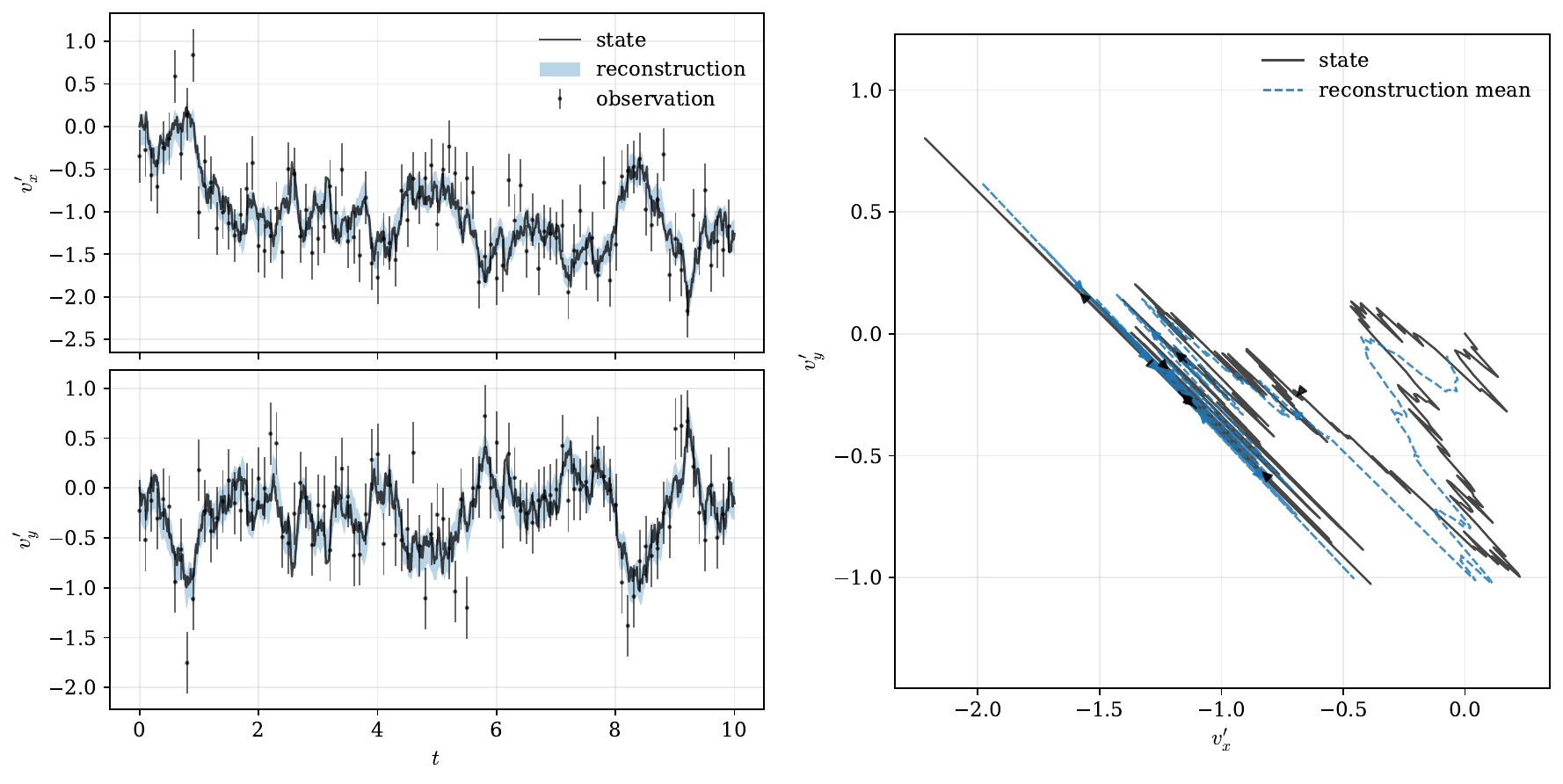}
\caption{Gaussian process reconstruction of the velocity of a particle exhibiting singular Brownian motion (state); with noisy measurements, state parameters $(\lambda_x, \sigma_y, f_x)=(1,1,-1)$, $\theta=45^\circ$, observation window $t \in [0, 10\tau_x]$, and zero initial conditions, where $\tau_x = \lambda_x^{-1}$. White Gaussian measurement noise $\boldsymbol{\varepsilon}_n \sim \mathcal{N}(\mathbf{0},\,\sigma_\varepsilon^2\mathbf{I}_2)$ is added to each observed state vector, with $\sigma_\varepsilon^2 = 0.1$. \textit{[left]} True state (black solid) and reconstructed state with 95\% credible intervals (blue bands). Black dots with error bars show observations. \textit{[right]} True state trajectory (black solid) and mean reconstructed trajectory (blue dashed). This is obtained using the fast Bayesian inference method described in Section~\ref{sec:fast_bayesian} \citep{2018PhRvE..98a2136S}.}
\label{fig:brownian_reconstruction}
\end{figure*}

\begin{figure*}
\centering
\includegraphics[width=0.6\textwidth]{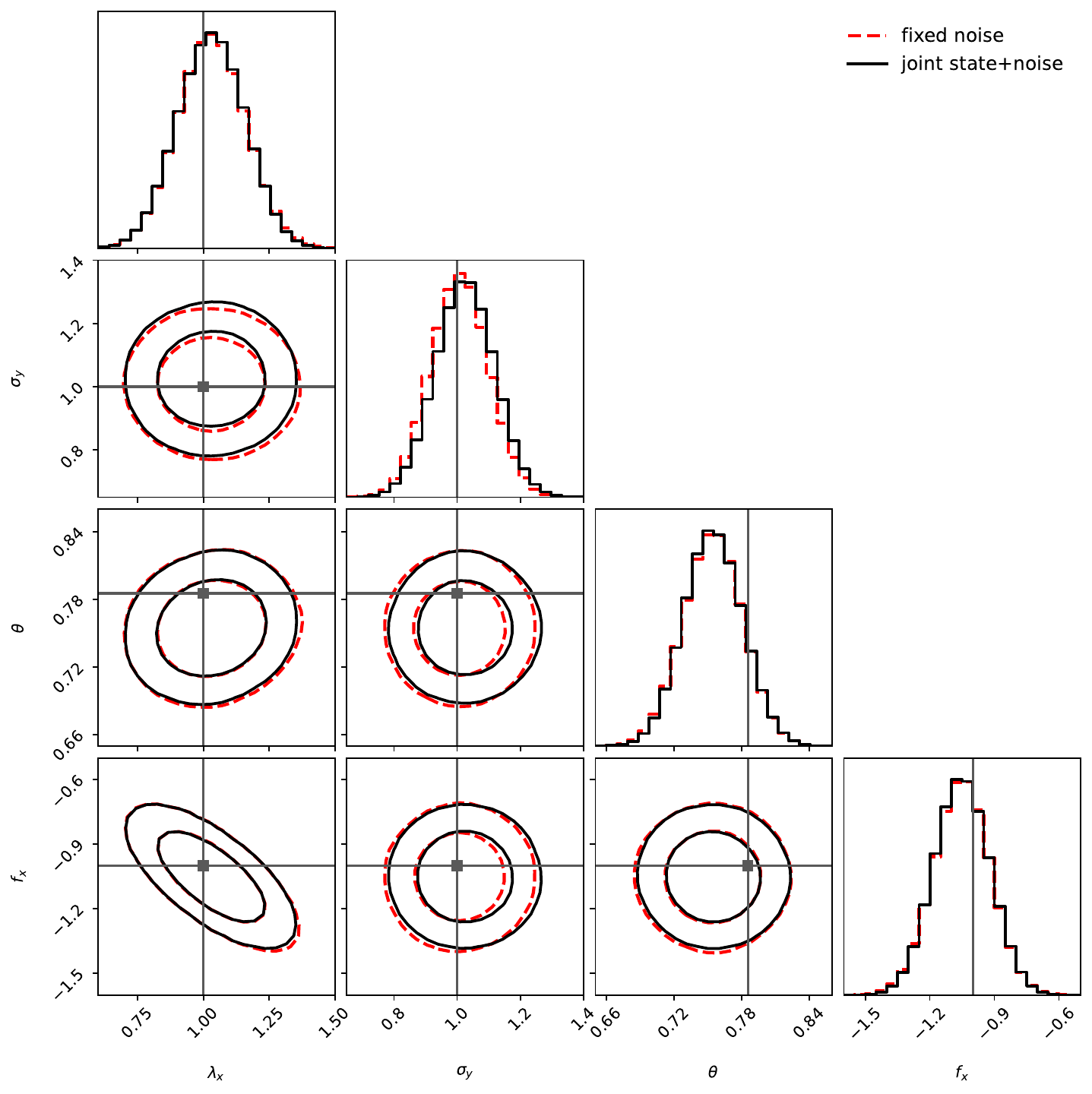}
\vspace{0.8em}
\includegraphics[width=0.8\textwidth]{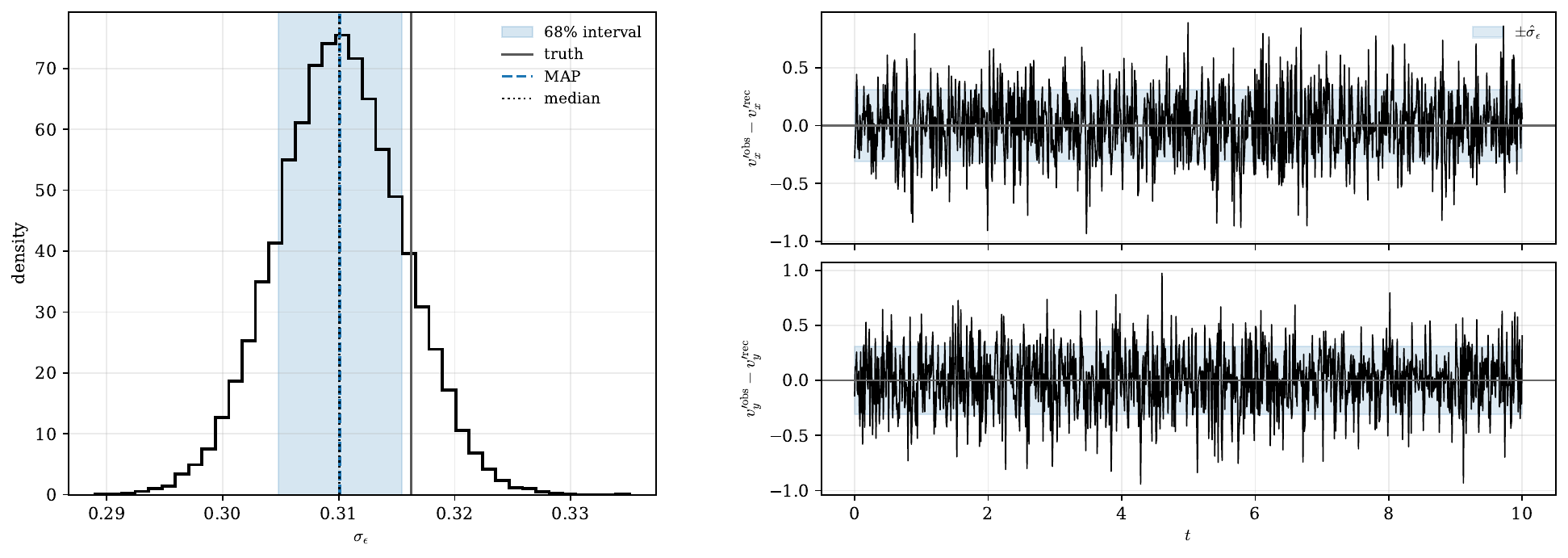}
\caption{
{\it Top panel}: Parameter posteriors (68\% and 95\% credible intervals) in singular Brownian motion (Section~\ref{sec:2d_brownian}) obtained using a fast Bayesian inference method \citep{2018PhRvE..98a2136S} with fixed noise (red dashed) and joint state \& noise estimation (black solid). True parameters are shown in gray: $(\lambda_x, \sigma_y, f_x)=(1,1,-1)$ and $\theta=45^\circ$. The state is sampled evenly ($N=1000$) with zero initial conditions and observation window $t \in [0, 10\tau_x]$, where $\tau_x = \lambda_x^{-1}$. The noisy measurements are obtained by adding white Gaussian noise $\boldsymbol{\varepsilon}_n \sim \mathcal{N}(\mathbf{0},\,\sigma_\varepsilon^2\mathbf{I}_2)$
to the state trajectory, with $\sigma_\varepsilon^2 = 0.1$.
{\it Bottom panel}: noise diagnostics for the residuals of the state reconstruction with joint state \& noise estimation. The left plot shows the measurement noise posterior together with the true value and the MAP estimate; right plot the whitened velocity residuals obtained after subtracting the reconstructed state out of the measurements and dividing by the reconstructed noise standard deviation.
}
\label{fig:brownian_posterior}
\end{figure*}

The OU process is Brownian motion \citep{Uhlenbeck:1930zz}. The result that the stochastic component of the phenomenological two-component model with spin wandering is an OU process suggests that we can tease out the body-averaged dynamics in a neutron star using Brownian motion as a toy model. For this reason we consider 2D Brownian motion with a singular feature that connects with the two-component model. We will also use the toy model to illustrate the fast Bayesian inference method \citep{2018PhRvE..98a2136S} for a regularized singular OU process.

Consider the eigenbasis of motion to be the $x$ and $y$ axes. Then, the equations of motion are given by
\begin{align}
    \dot{v}_x(t) &= -\lambda_x v_x(t) + f_x + \sqrt{Q_x} \dot{W}_x(t) \,, \\
    \dot{v}_y(t) &= -\lambda_y v_y(t) + f_y + \sqrt{Q_y} \dot{W}_y(t) \,, \\
    \langle \dot{W}_i(t) \dot{W}_j(t') \rangle &= \delta_{ij} \delta(t - t') \,,
\end{align}
where $\lambda_x=\tau_x^{-1}$ and $\lambda_y=\tau_y^{-1}$ are relaxation rates along the $x$ and $y$ directions, respectively, $f_x$ and $f_y$ are constant forces per unit mass acting on the particle in the $x$ and $y$ directions, respectively, and $\dot{W}_x(t)$ and $\dot{W}_y(t)$ are independent Gaussian white noise processes. The constants $Q_x$ and $Q_y$ are diffusion coefficients in the $x$ and $y$ directions, respectively. The above equations describe a shifted 2D OU process,
\begin{equation}
    \dot{\mathbf{v}}(t) = -\boldsymbol{\lambda} \mathbf{v}(t) + \mathbf{f} + \boldsymbol{\sigma} \dot{\mathbf{W}}(t) \,,
\end{equation}
with diagonal drift and volatility matrices:
\begin{equation}
    \boldsymbol{\lambda} = \begin{bmatrix} \lambda_x & 0 \\ 0 & \lambda_y \end{bmatrix} \,, \quad
    \boldsymbol{\sigma} = \begin{bmatrix} \sqrt{Q_x} & 0 \\ 0 & \sqrt{Q_y} \end{bmatrix} \,,
\end{equation}

The observation basis is the $x'$ and $y'$ axes; which are rotated by an angle $\theta$ with respect to the $x$ and $y$ axes. The equations of motion in this frame can be derived by appropriately applying the rotation matrix
\begin{equation}
    \mathbf{R} = \begin{bmatrix} \cos\theta & -\sin\theta \\ \sin\theta & \cos\theta \end{bmatrix} 
\end{equation}
to the vectors and matrices in the eigenbasis; i.e., vectors and matrices in the observation basis are given by
\begin{align}
    \mathbf{v}' &= \mathbf{R} \mathbf{v} \,, \\
    \mathbf{f}' &= \mathbf{R} \mathbf{f} \,, \\
    \boldsymbol{\lambda}' &= \mathbf{R} \boldsymbol{\lambda} \mathbf{R}^\top \,, \\
    \boldsymbol{\sigma}' &= \mathbf{R} \boldsymbol{\sigma} \mathbf{R}^\top \,,
\end{align}
and the equations of motion given by
\begin{equation}
    \dot{\mathbf{v}}'(t) = -\boldsymbol{\lambda}' \mathbf{v}'(t) + \mathbf{f}' + \boldsymbol{\sigma}' \dot{\mathbf{W}}(t) \,.
\end{equation}
In contrast with the eigenbasis, the drift and volatility matrices in the observation basis are non-diagonal and the processes along the $x'$ and $y'$ axes are correlated. The constant force per unit mass $\mathbf{f}$ provides a long time deterministic (linear) trend $\mathbf{v}'_{\rm D}$ in the observed velocity $\mathbf{v}'$. The velocity residuals $\mathbf{v}'_{\rm S}=\mathbf{v}'-\mathbf{v}'_{\rm D}$ are fully specified by the drift and volatility matrices.

For illustration we consider a special case: $\lambda_y=0, f_y=0, \sigma_x=0$. This turns the problem into a singular OU process because one of the eigenvalues of the drift matrix is zero. Then, the process along the $y$ axis is diffusive and the process along the $x$ axis is OU for $\lambda_x > 0$. The deterministic part is parametrized in the eigenbasis as $v_x(t) = v_{x,0} e^{-\lambda_x t} + f_x (1 - e^{-\lambda_x t}) / \lambda_x$; which at long times compared to the relaxation time $\tau_x$ approaches the steady state value $\llangle v_x \rrangle = f_x / \lambda_x$. The observed processes along the $x'$ and $y'$ axes are a superposition of the OU and diffusive eigenbasis processes; with amounts determined by the rotation angle $\theta$. The deterministic drift in the observation basis can be obtained by applying the rotation matrix to $\mathbf{v}_{\rm D} = (\llangle v_x \rrangle, 0)^\top$; this gives $\mathbf{v}'_{\rm D} = ( \llangle v_x \rrangle \cos \theta, \llangle v_x \rrangle \sin \theta )^\top$. This captures key features of the two-component model with spin wandering: one velocity eigenmode has a finite relaxation time and the other an infinite relaxation time. The external forces support a deceleration. Realizations of the process are shown in Figure \ref{fig:brownian_realizations}. This is reminiscent of realizations of the crust and superfluid angular velocities in the two-component model with spin wandering \citep{Antonelli:2022gqw}.

Figures \ref{fig:brownian_reconstruction}--\ref{fig:brownian_posterior} illustrate the application of the fast Bayesian inference method \citep{2018PhRvE..98a2136S} for a singular OU process; or Brownian motion with diffusive and damped eigenmodes. The method was applied to reconstruct the velocity of a particle exhibiting random Brownian motion and to infer the parameters of the system with clean and noisy measurements. We consider the state $(\lambda_x, \sigma_y, f_x)=(1,1,-1)$ and $\theta=45^\circ$. Measurement noise was accommodated by adding white Gaussian noise $\boldsymbol{\varepsilon}_n \sim \mathcal{N}(\mathbf{0},\,\sigma_\varepsilon^2\mathbf{I}_2)$ to the state trajectory with $\sigma_\varepsilon^2 = 0.1$. The choices are arbitrary for illustration of the effects of different forces acting on the system (long term drift to $\llangle v_x \rrangle = f_x/\lambda_x$, OU in $v_x$ and diffusion in $v_y$). The physics holds for arbitrary values. The path reconstruction (Figure \ref{fig:brownian_reconstruction}) shows that the state is consistently tracked by the reconstruction. The parameter posteriors and noise diagnostics (Figure \ref{fig:brownian_posterior}) show that the parameters can be inferred with noisy measurements.

\section{Analytic semiseparable likelihood for full-state angular-velocity observations}
\label{sec:full_state_semiseparable}

In Section~\ref{subsec:analytic_semiseparable} we specialize to the
case that is relevant to radio pulsar timing, where only the crust
angular velocity is observed.  For completeness, we write down the
full-state construction based on the time-domain analytic solutions
in \cite{Bernardo:2026opv} for the case in which both angular-velocity
components are directly observed: the crust angular velocity
$\Omega_{\rm c}$ by radio timing and the superfluid angular velocity
$\Omega_{\rm s}$ by, say, continuous gravitational-wave observations.
The crust-only likelihood of the main text follows as a special case.

At each epoch $t_i$ the observation is a two-component vector,
\begin{equation}
    {\bf y}_i
    =
    \begin{bmatrix}
        \delta\Omega_{\rm c}(t_i) \\
        \delta\Omega_{\rm s}(t_i)
    \end{bmatrix}
    +
    {\bf M}_i\boldsymbol{\beta}
    +
    \boldsymbol{\varepsilon}_i,
    \label{eq:full_state_obs}
\end{equation}
where ${\bf M}_i$ is a $2\times p$ deterministic trend matrix,
$\boldsymbol{\beta}$ is a $p$-vector of nuisance trend parameters,
and $\boldsymbol{\varepsilon}_i$ is heteroscedastic measurement noise
with
\begin{equation}
    \langle
        \boldsymbol{\varepsilon}_i
        \boldsymbol{\varepsilon}_j^\top
    \rangle
    =
    \boldsymbol{\Sigma}_i\delta_{ij},
    \qquad
    \boldsymbol{\Sigma}_i
    =
    \begin{bmatrix}
        \sigma_{{\rm c},i}^2 & 0 \\
        0 & \sigma_{{\rm s},i}^2
    \end{bmatrix}.
\end{equation}
The covariance matrix between epochs $t_i$
and $t_j$ is
\begin{equation}
    {\bf C}_{ij}(\boldsymbol{\theta})
    =
    {\bf K}(t_i,t_j;\boldsymbol{\theta})
    +
    \boldsymbol{\Sigma}_i\delta_{ij},
    \label{eq:full_state_cov}
\end{equation}
where
\begin{equation}
    {\bf K}(t,t')
    =
    \begin{bmatrix}
        K_{\rm cc}(t,t') & K_{\rm cs}(t,t') \\
        K_{\rm cs}(t',t) & K_{\rm ss}(t,t')
    \end{bmatrix},
    \label{eq:full_state_K}
\end{equation}
with $K_{\rm cc}$, $K_{\rm ss}$, and $K_{\rm cs}$ given by the
finite-start expressions derived in \cite{Bernardo:2026opv}.

Setting $x=t-t_0$ and $x'=t'-t_0$, each entry of ${\bf K}$
is a bilinear combination of the three scalar mode kernels,
\begin{align}
    K_+(t,t') &= Q_+\min(x,x'),\\
    K_-(t,t') &=
    \frac{Q_-\tau}{2}
    \left[
        e^{-|x-x'|/\tau}
        -
        e^{-(x+x')/\tau}
    \right],\\
    K_{+-}(t,t') &=
    Q_\times\tau
    \left[
        e^{-(x'-x)/\tau}\Theta(x'-x)
        +\Theta(x-x')
        -e^{-x'/\tau}
    \right],
\end{align}
where $K_{+-}(t,t')=\langle\Delta\Omega_+(t)\Delta\Omega_-(t')
\rangle$.  In terms of the weights
\begin{equation}
    w_- = \frac{\tau_{\rm s}}{\tau_{\rm c}+\tau_{\rm s}},
    \qquad
    v_- = \frac{\tau_{\rm c}}{\tau_{\rm c}+\tau_{\rm s}},
    \qquad
    w_+ = \frac{\tau}{\tau_{\rm c}+\tau_{\rm s}},
\end{equation}
the three entries of ${\bf K}$ are
\begin{align}
    K_{\rm cc}(t,t') &=
    w_-^2 K_-(t,t')
    + w_-w_+\left[K_{+-}(t,t')+K_{+-}(t',t)\right]
    \nonumber\\&\quad
    + w_+^2 K_+(t,t'),
    \label{eq:Kcc}\\
    K_{\rm ss}(t,t') &=
    v_-^2 K_-(t,t')
    - v_-w_+\left[K_{+-}(t,t')+K_{+-}(t',t)\right]
    \nonumber\\&\quad
    + w_+^2 K_+(t,t'),
    \label{eq:Kss}\\
    K_{\rm cs}(t,t') &=
    -w_-v_- K_-(t,t')
    + w_+\left[w_- K_{+-}(t',t)-v_- K_{+-}(t,t')\right]
    \nonumber\\&\quad
    + w_+^2 K_+(t,t').
    \label{eq:Kcs}
\end{align}

For $i>j$, each block ${\bf K}(t_i,t_j)$ factorizes as
\begin{equation}
    {\bf K}(t_i,t_j)
    =
    {\bf F}_i{\bf G}_j^\top,
    \qquad i>j,
    \label{eq:full_state_semisep}
\end{equation}
where ${\bf F}_i$ and ${\bf G}_j$ are $2\times 2$ factor matrices.
Using the shorthand $s_i=e^{-x_i/\tau}$, both rows of ${\bf F}_i$
share the same two scalar basis functions,
\begin{equation}
    {\bf F}_i
    =
    \begin{bmatrix}
        w_-\, s_i & w_+ \\
        -v_-\, s_i & w_+
    \end{bmatrix},
    \label{eq:full_state_F}
\end{equation}
and the coefficient matrix has row-specific entries reflecting
the asymmetric coupling of the two eigenmodes to the crust and
superfluid components. We introduce the shorthand
\begin{align}
    B_j &= \frac{Q_-\tau}{2}
    \left(e^{x_j/\tau}-e^{-x_j/\tau}\right),
    \label{eq:Bj_def}\\
    A_j &= Q_\times\tau
    \left(e^{x_j/\tau}-1\right),
    \label{eq:Aj_def}\\
    D_j &= Q_\times\tau
    \left(1-e^{-x_j/\tau}\right),
    \label{eq:Dj_def}\\
    C_j &= Q_+\,x_j,
    \label{eq:Cj_def}
\end{align}
where $B_j$ is the coefficient of the damped-mode covariance
$K_-$, $C_j$ is the coefficient of the diffusive-mode covariance
$K_+$, $A_j$ is the coefficient associated with
$K_{+-}(t_j,t_i)$, and $D_j$ is the coefficient associated with
$K_{+-}(t_i,t_j)$ for $i>j$. The coefficient matrix is then
\begin{equation}
    {\bf G}_j
    =
    \begin{bmatrix}
        w_- B_j + w_+ A_j
        &
        w_- D_j + w_+ C_j
        \\
        -v_- B_j + w_+ A_j
        &
        -v_- D_j + w_+ C_j
    \end{bmatrix},
    \label{eq:full_state_G}
\end{equation}
so that the crust row of ${\bf G}_j$ is
\begin{align}
    G_j^{{\rm c},1} &= w_- B_j + w_+ A_j,
    \label{eq:Gc1}\\
    G_j^{{\rm c},2} &= w_- D_j + w_+ C_j,
    \label{eq:Gc2}
\end{align}
and the superfluid row is
\begin{align}
    G_j^{{\rm s},1} &= -v_- B_j + w_+ A_j,
    \label{eq:Gs1}\\
    G_j^{{\rm s},2} &= -v_- D_j + w_+ C_j.
    \label{eq:Gs2}
\end{align}
One can verify directly that ${\bf F}_i{\bf G}_j^\top$ reproduces
all three entries of ${\bf K}(t_i,t_j)$ for $i>j$:
\begin{align}
    \left[{\bf F}_i{\bf G}_j^\top\right]_{11}
    &= w_- s_i\,G_j^{{\rm c},1}+w_+ G_j^{{\rm c},2}
    = K_{\rm cc}(t_i,t_j),
    \label{eq:check11}\\
    \left[{\bf F}_i{\bf G}_j^\top\right]_{22}
    &= -v_- s_i\,G_j^{{\rm s},1}+w_+ G_j^{{\rm s},2}
    = K_{\rm ss}(t_i,t_j),
    \label{eq:check22}\\
    \left[{\bf F}_i{\bf G}_j^\top\right]_{12}
    &= w_- s_i\,G_j^{{\rm s},1}+w_+ G_j^{{\rm s},2}
    = K_{\rm cs}(t_i,t_j).
    \label{eq:check12}
\end{align}
The crust row coincides with the scalar factors
$[V_j^{(1)},V_j^{(2)}]$ of the crust-only likelihood in
Section~\ref{subsec:analytic_semiseparable} upon identifying
$V_j^{(1)}=G_j^{{\rm c},1}$ and $V_j^{(2)}=G_j^{{\rm c},2}$,
confirming that the crust-only construction is the exact
reduction of the full-state one.  The superfluid row
$G_j^{{\rm s},\cdot}$ differs from the crust row because the
lag mode $\Omega_-$ couples to the superfluid with weight $-v_-$
rather than $w_-$, while the diffusive mode $\Omega_+$ retains
the same weight $w_+$ in both components.  The rank of the
semiseparable structure is still governed by $J=2$ scalar basis
functions $\{s_i,\,1\}$; the row-specific coefficients in
${\bf G}_j$ reflect the asymmetric coupling of the two eigenmodes
to the two components and do not increase the rank.
The diagonal blocks are stored separately,
\begin{equation}
    {\bf A}_i
    =
    {\bf K}(t_i,t_i)
    +
    \boldsymbol{\Sigma}_i,
    \label{eq:full_state_diag_block}
\end{equation}
and the upper triangle follows by
${\bf K}(t_i,t_j)={\bf K}(t_j,t_i)^\top$.

\subsection*{Trend marginalization and posterior}

With the stacked data vector
${\bf y}=({\bf y}_1^\top,\ldots,{\bf y}_N^\top)^\top$ and the
stacked trend matrix ${\bf M}$, the generalized least-squares
trend estimate at fixed $\boldsymbol{\theta}$ is
\begin{equation}
    \hat{\boldsymbol{\beta}}
    =
    \left({\bf M}^\top{\bf C}^{-1}{\bf M}\right)^{-1}
    {\bf M}^\top{\bf C}^{-1}{\bf y}.
\end{equation}
With ${\bf r}={\bf y}-{\bf M}\hat{\boldsymbol{\beta}}$, the
trend-marginalized likelihood is
\begin{align}
    \ln{\cal L}(\boldsymbol{\theta})
    =
    -\frac{1}{2}
    \bigg[
        {\bf r}^\top{\bf C}^{-1}{\bf r}
        &+\ln|{\bf C}|
        \nonumber\\
        &+\ln\!\left|{\bf M}^\top{\bf C}^{-1}{\bf M}\right|
        +(2N-p)\ln(2\pi)
    \bigg].
    \label{eq:full_state_trend_marg_likelihood}
\end{align}
The block semiseparable Cholesky recursion of
Section~\ref{subsec:analytic_semiseparable} generalizes to $2\times 2$
blocks by replacing scalars with $2\times 2$ matrices throughout;
the scalar pivots $d_i$ become $2\times 2$ pivot matrices ${\bf D}_i$,
and the log-determinant accumulates as
$\ln|{\bf C}|=\sum_i\ln|{\bf D}_i|$.
The factorization costs ${\cal O}(N)$ with a fixed prefactor from
the $2\times 2$ arithmetic.

\subsection*{Reduction and hidden-state reconstruction}

The crust-only likelihood of Section~\ref{subsec:analytic_semiseparable}
is recovered by retaining only the first row of ${\bf F}_i$ and
the first row of ${\bf G}_j$, which gives back the scalar factors
${\bf U}_i=[w_-s_i,\,w_+]$ and ${\bf V}_j=[V_j^{(1)},V_j^{(2)}]$
of Eq.~\eqref{eq:semisep_Kc}.

When only the crust is observed, the hidden superfluid residual
can be reconstructed by Gaussian conditioning.  At fixed
$\boldsymbol{\theta}$, the conditional mean and variance of the
superfluid residual at epoch $t_\star$ are
\begin{equation}
    \left\langle
        \delta\Omega_{\rm s}(t_\star)
        \,\middle|\,
        {\bf r}
    \right\rangle
    =
    {\bf k}_{\rm cs}(t_\star)^\top
    {\bf C}_{\rm cc}^{-1}{\bf r},
    \label{eq:superfluid_reconstruction_mean}
\end{equation}
\begin{equation}
    {\rm Var}\!\left[
        \delta\Omega_{\rm s}(t_\star)
        \,\middle|\,
        {\bf r}
    \right]
    =
    K_{\rm ss}(t_\star,t_\star)
    -
    {\bf k}_{\rm cs}(t_\star)^\top
    {\bf C}_{\rm cc}^{-1}
    {\bf k}_{\rm cs}(t_\star),
    \label{eq:superfluid_reconstruction_var}
\end{equation}
where ${\bf C}_{\rm cc}$ is the $N\times N$ crust-only covariance
matrix and $[{\bf k}_{\rm cs}(t_\star)]_i=K_{\rm cs}(t_i,t_\star)$.
Both inverse-covariance actions use the crust-only rank-two
semiseparable solve and cost ${\cal O}(N)$.


\label{lastpage}
\end{document}